\documentclass[12pt,draftclsnofoot,onecolumn]{IEEEtran}
\IEEEoverridecommandlockouts

\usepackage{graphicx}
\usepackage{algorithm}
\usepackage{algorithmic}

\usepackage[bottom]{footmisc}
\usepackage{subcaption}
\usepackage{cite}
\usepackage{url}
\usepackage{mdwmath}     
\usepackage{mdwtab}
\usepackage{amsfonts,amsmath,color,amssymb,amsxtra,amsbsy,floatflt}
\usepackage{indentfirst} 
\usepackage{tabularx}
\usepackage{booktabs}
\usepackage{multirow}
\usepackage{nth}
\usepackage{comment}
\usepackage{epstopdf}
\usepackage{bbold}
\usepackage{multirow}
\usepackage{hhline}
\usepackage{multicol}

\graphicspath{ {figures/} }

\newlength{\oldtextfloatsep}

\DeclareMathOperator*{\argmax}{arg\,max}

\def\1{\mathbf{1}}
\newcommand{\vv}[1]{\boldsymbol{#1}}
\newcommand{\vvi} [2]{ \boldsymbol{#1}^{(#2)} }

\newcounter{subeqn} %
\makeatletter
\@addtoreset{subeqn}{equation}
\makeatother

\newcounter{subproblem}

\newcommand{\name}{LACO}
\newcommand{\change}[1]{{\color{black}{#1}}}

\begin{document}
\title{\name: A Latency-Driven Network Slicing Orchestration in Beyond-5G Networks}

\author{Lanfranco Zanzi,~\IEEEmembership{Student Member,~IEEE,}\\
	    Vincenzo~Sciancalepore,~\IEEEmembership{Senior Member,~IEEE,}
	    Andres Garcia-Saavedra,
	    Hans~D.~Schotten,~\IEEEmembership{Member,~IEEE,}
	    and~Xavier~Costa-P\'erez,~\IEEEmembership{Senior Member,~IEEE}
\thanks{\textit{L. Zanzi and Hans D. Schotten are with Institute of Wireless Communication, Technische Universit\"at Kaiserslautern, 67655 Kaiserslautern, Germany. Emails: schotten@eit.uni-kl.de}}%
	\thanks{\textit{L. Zanzi, V. Sciancalepore, A. Garcia-Saavedra and X. Costa-P\'erez are with NEC Laboratories Europe GmbH., 69115 Heidelberg, Germany. Emails: \{lanfranco.zanzi, vincenzo.sciancalepore, andres.garcia.saavedra, xavier.costa\}@neclab.eu}}%
}

\maketitle

\begin{abstract}

\emph{Network Slicing} is expected to become a game changer in the upcoming 5G networks and beyond, enlarging the telecom business ecosystem through still-unexplored vertical industry profits.
This implies that heterogeneous service level agreements (SLAs) must be guaranteed \emph{per slice} given the multitude of predefined requirements. 

In this paper, we pioneer a novel radio slicing orchestration solution that simultaneously provides \emph{latency} and {throughput} guarantees in a multi-tenancy environment. Leveraging on a solid mathematical framework, we exploit the exploration-vs-exploitation paradigm by means of a multi-armed-bandit-based (MAB) orchestrator, \name, that makes adaptive resource slicing decisions with no prior knowledge on the traffic demand or channel quality statistics. As opposed to traditional MAB methods that are blind to the underlying system, \name{} relies on system structure information to expedite decisions. After a preliminary simulations campaign empirically proving the validness of our solution, \change{we provide a robust implementation of \name{} using off-the-shelf equipment to fully emulate realistic network conditions: near-optimal results within affordable computational time are measured when \name{} is in place.}
\end{abstract}


\IEEEpeerreviewmaketitle

\thispagestyle{empty}	

\section{Introduction}
\label{intro}
\begin{figure}[t!]
      \centering
      \includegraphics[trim = 2cm 5cm 5cm 1cm, clip, width=0.8\linewidth ]{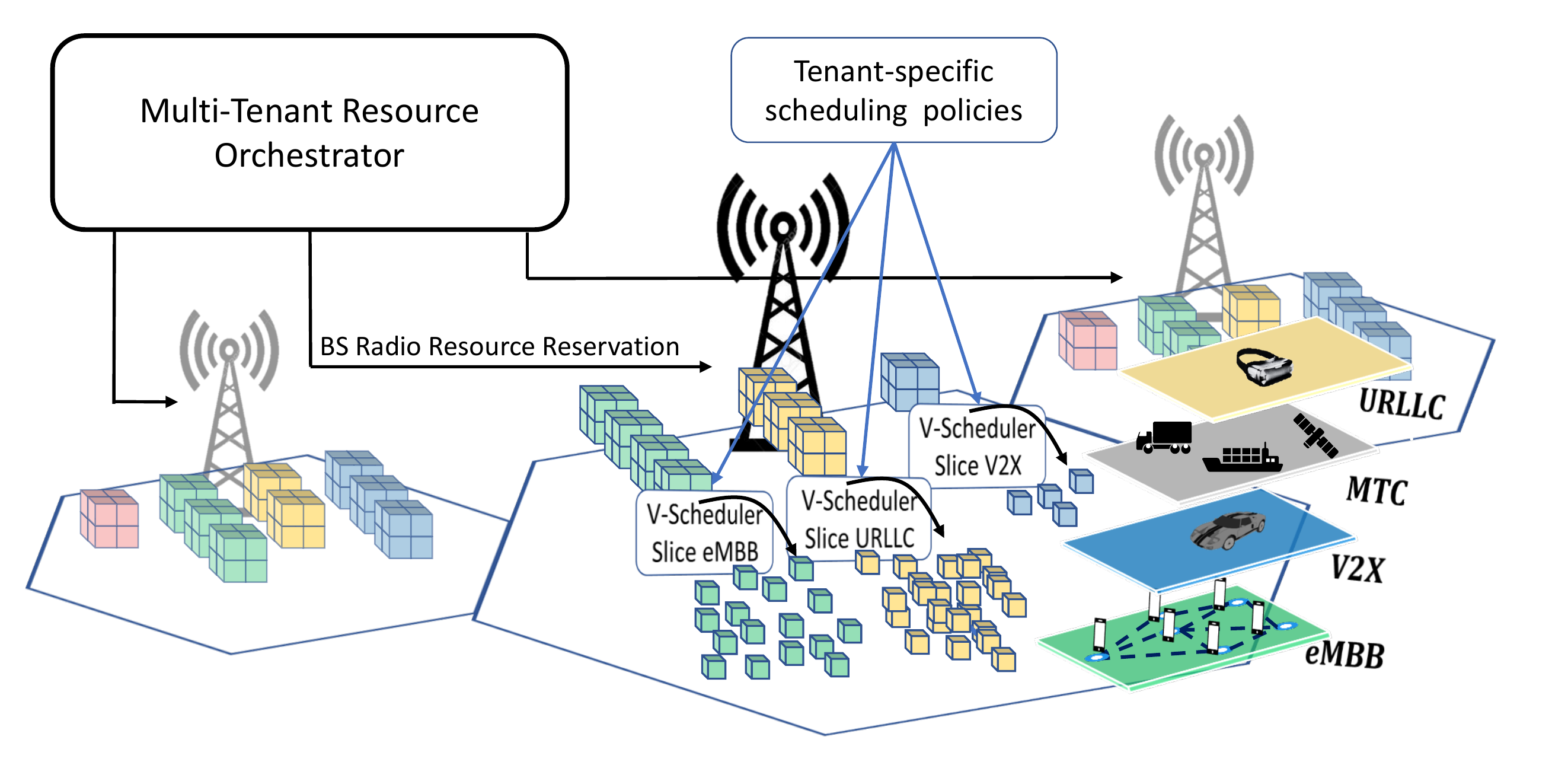}
      \vspace{-7mm}
      \caption{\change{ \small Illustration of the network slicing concept.}}
      \label{fig:slicing}
\end{figure}

The quest for new sources of revenue that revitalizes the mobile industry has spawned an unprecedented hype around the fifth-generation of mobile networks (5G) and, in particular, the network slicing concept. Enabled by software-defined networking (SDN) and network function virtualization (NFV), network slicing allows telco operators to offer virtualized slices of infrastructure resources \emph{on-demand} to heterogeneous 3${^{\text{rd}}}$-party services~\cite{WhitePaperMWC}. 
A high-level view of the system considered in this paper is described in Fig.~\ref{fig:slicing}. The figure represents a series of \emph{sliceable} base stations as a pool of radio resources (coloured cubes in the figure). The resource allocation process is considered hierarchical: while \emph{bundles} of radio resources are assigned to different tenants (namely radio slices), each tenant autonomously schedules its bundle of radio resources to each individual user following classic radio scheduling policies. The difference between such operations is subtle but of paramount importance: a slice controller operates at a larger timescale and thus over a coarser granularity~\cite{Slicing2018CoNEXT, foukas_orion}. \change{While most prior work on network slicing focuses on average bit-rate guarantees~\cite{foukas_orion, Sciancalepore2017Optimising}, \emph{latency} considerations have received little attention. Latency aspects however are gaining more and more attraction as a quest to face advanced use-cases requirements, e.g., autonomous driving and platooning~\cite{Lee2018_lowlatency} in Vehicle-to-everything (V2X) enabled scenarios. In this context, accurate resource allocation schemes and inter-slice isolation aspects are fundamental features to support the provisioning of latency-constrained services. }

Given the plethora of works on classic radio scheduling~\cite{Sharma,Lee}, we keep this aspect out of the scope of this paper and we focus instead on the former impelling need: a proper design of an orchestration solution that autonomously assigns chunks of radio spectrum (slices) in relatively larger time-scales pursuing the goal of \emph{guaranteeing simultaneously latency and throughput constraints}. From the best of our knowledge, there is a non-negligible lack of works focusing on both aspects simultaneously in sliced-network environments.

To fill this gap, we design a LAtency-Controlled Orchestrator (\name{}), a network slice controller that maps virtual radio resource allocations to physical resources while still guaranteeing latency requirements\footnote{Note that \name{} does \emph{not} compete with state-of-the-art throughput-only slice controllers---in fact, we purposely assume the presence of an admission controller that ensures that the aggregate load incurred by granted slices is within the system capacity region.}. Specifically, \name{} augments such prior work by accommodating resources to (granted) slices such that latency agreements are satisfied. This unlocks a new business opportunity for the telco operators
that may apply customized pricing models according to the \emph{elasticity} of offered slice latency constraints.

{\bf Technical challenges.} While designing \name{}, two sources of uncertainty need to be under control: $i$) the behavioral dynamics of the (aggregated) demand across involved tenants and $ii$) the inherent randomness of the wireless channel. These system dynamics have been traditionally modeled via either complex solutions that are hard to solve in realistic settings or via simplistic assumptions at the expense of low performance figures. In our work, we explore a novel approach by designing a scheme that \emph{learns} the implications that allocation decisions have on per-slice latency without explicitly making assumptions on the underlying dynamics. To this aim, we first model our decision-making problem as a Markov Decision Process\footnote{With a little misuse of nomenclature, we will refer to Markov Decision Process (MDP) rather than Semi-Markov Decision Process (SMDP) despite considering continuous time scales.} (MDP), which allows us to neglect low-level details of the tenant demands and channel dynamics while letting us retain some knowledge on the consequences that a given action may have on the most immediate next system state. 

An MDP model helps us to fully explore the problem features. However, the process of learning the state transition probability matrix of each of the embedded Markov chains incurs in prohibitive overhead as a reinforcement learning agent has to explore the whole space of state-action trajectories---the so-called \emph{curse of dimensionality}. To address this, we resort to a Multi-Armed Bandit (MAB) model where the attained reward depends only on the action taken from a bounded set of possible actions. Importantly, in contrast to traditional MAB methods, \name{} is \emph{model-aware} (though not model-dependant), i.e., it exploits (abstracted) information regarding the underlying system to expedite the selection of highly rewarding actions, which is particularly attractive when dealing with dynamic non-stationary scenarios. 

The main contributions of our paper can be summarized as follows:
\begin{itemize}
    \item We introduce a Discrete-Time Markov Chain (DTMC) model to capture the dynamics of the (instantaneous) aggregate slice traffic demand and the wireless channel variations.
    \item We present a latent variable regression model to accurately anticipate the transition probability matrix of the proposed DTMCs.
    \item We formulate the dynamic slice resource provisioning as a Markov Decision Process (MDP).
    \item We design a \emph{model-aware} Multi-Armed Bandit (MAB) method to guide the decision-making process, which relies on the above DTMC models and anticipated transition probabilities to speed up convergence.
    \item We present an exhaustive simulations campaign to assess the performance of our approach.
    \item We implement and field-test our solution using off-the-shelf equipment that emulates real network conditions: \name{} shows its innovative performance gain against considered legacy techniques.
\end{itemize}

The remainder of the paper is structured as follows. Section~\ref{sect:framework} formulates our problem and presents the main building blocks of \name{}. Section~\ref{sect:MC} introduces an DTMC model that helps us expedite the action-space exploration phase and Section~\ref{sect:monitor} deeply analyzes it. In Section~\ref{sect:mdp}, we introduce our decision process as a Markov Decision Process (MDP) and present a model-aware Multi-Armed Bandit decision-making engine integrated in \name{}. Section~\ref{sect:perf_eval} presents our preliminary simulation campaign to validate the design principles of \name{}, whereas Section~\ref{sect:poc} details the implementation of our novel solution into off-the-shelf equipment with realistic network performance. Finally, Section~\ref{sect:related} summarizes related literature and Section~\ref{sect:concl} concludes the paper with some final remarks.

\section{\name: The framework overview}
\label{sect:framework}
 \change{ Our solution relies on the concept of slicing-enabled networks wherein multiple network tenants are willing to obtain a network slice with predefined service level agreements (SLAs). Such SLAs may be expressed in terms of maximum slice throughput and average access latency. Within the context of our paper, we define the average access latency as time the traffic belonging to a certain slice needs to wait before being served due to scheduling procedures.} In particular, we focus on the radio access network (RAN) domain and design \name, a RAN controller that dynamically provisions spectrum resources to admitted network slices while providing latency guarantees. In the following, we overview the main system building blocks with detailed notation and assumptions.

\subsection{Business scenario}
We consider different entities in our system: $i$) an \emph{infrastructure provider} owning the physical infrastructure who offers isolated  RAN slices as a service, $ii$) \emph{tenants} who acquire and manage slices with given SLAs to deliver services to end-users, and $iii$) \emph{end-users}, who demand radio resources from such tenants/slices. 

Let us define $\mathcal{I}$ as the set of running network slices and $\mathcal{U}_i$ as the set of end-users associated to the $i$-th slice. \change{The total amount of wireless resources (radio spectrum) is split into} multiple non-overlapping network slices, each one belonging to one single tenant $i\in\mathcal{I}$.\footnote{We assume a one-to-one mapping between slices and tenants. Therefore, we use $i\in\mathcal{I}$ interchangeably throughout the paper as a tenant identifier or its associated slice. \change{Note that this assumption can be easily relaxed in the model.}} Based on fixed SLAs, each network slice is characterized by~\emph{maximum throughput} and \emph{expected latency} denoted by $\Lambda_i$ and $\Delta_i$, respectively. We assume that an admission control process\footnote{Given the plethora of solutions in the literature, the admission control design is out of the scope of this work. We refer the reader, for example, to~\cite{Slicing2018CoNEXT, Sciancalepore2017Optimising} for more details.} \change{is concurrently running on a higher tier so that the average aggregate load can be accommodated within the overall system capacity.}

\subsection{Notation} 
We use conventional notation. We let $\mathbb{R}$ and $\mathbb{Z}$ denote the set of real and integer numbers. 
We use $\mathbb{R}_+$, $\mathbb{R}^n$, and $\mathbb{R}^{n\times m}$ to represent the sets of non-negative real numbers, $n$-dimensional real vectors, and $m\times n$ real matrices, respectively. Vectors are denoted as column vectors and written in bold font. Subscripts represent an element in a vector and superscripts elements in a sequence. For instance, $\langle \vvi x{t} \rangle$ is a sequence of vectors with $\vvi x{t} = [ x^{(t)}_1, \dots,  x^{(t)}_n   ]^T$ being a vector from $\mathbb{R}^n$, and  $x^{(t)}_i$ the $i$'th component of the $t$'th vector in the sequence. Operation $[\cdot]^T$ represents the transpose operator while $[x_1,\dots,x_n]_{\text{diag}}$ translates the vector into a diagonal matrix. Last, $\vv{\mathbb{1}}$ and $\vv{\mathbb{0}}$ indicate an all-ones and all-zeroes vector, respectively, and $\lceil\cdot\rceil$ is the ceiling operation.

\begin{table}[!ht]
\footnotesize
\centering
\caption{Notation table}
\vspace{-4mm}
\label{tab:notations}
\begin{tabular}{|l|l||l|l||l|l|}
\hline
Notation & Description & Notation & Description & Notation & Description\\ \hline
$y_i$                 & Slice configuration      & $z_\sigma$ & Arm selection freq. & $\phi\in\Phi$   & Action index  \\
$n \in N$                   & Decision epoch index     & $\mathcal{N}_i(\mu_i ,\nu_i^2)$\!\! & Normal distribution & $\omega(\cdot)$ & Latent var. weight \\
$u_i\in\mathcal{U}_i$ & User index               & $m\in\mathcal{M}$ & MCS index & $R(\cdot)$ & Reward function    \\
$d\in\{0,1\}$         & Exceed delay flag        & $i\in\mathcal{I}$ & Slice index & $\psi(\sigma)$  & Accuracy value     \\
$\lambda_i^{(n)}$     & Inst. traffic demand     & $r_i^{(n)}$ & Bits not served & $L_i$ & Latency constraint \\
$f(x,\rho_i)$         & Traffic demands distr.   & $\bar{u}_i$ & Aggregate user & $\Gamma_m$ & Bits per subframe  \\
$\zeta(\cdot)^{(n)}$  & Throughput mapping       & $g\in\mathcal{G}$ & Channel level & $f(x,\theta_i)$ & Channel distr.     \\
$\sigma\in\Sigma$     & MDP state index          & $\gamma_i^{(n)}$ & Inst. SNR & $\Delta_i$ & Latency tolerance  \\
$w\in\mathcal{W}$     & (Latent) Channel quality & \change{$C$} & \change{Capacity of BS} & \change{$\tau$} & \change{Rayleigh scale param.}  \\
$T(\cdot)$    & Transition function     & \change{$\epsilon$} & \change{Decision interval duration} & \change{$\Theta$} & \change{PRB chunk} \\
\hline

\end{tabular}
\vspace{-8mm}
\end{table}

\subsection{Problem Definition}
\label{sect:prob_def}
\change{
Assuming that an instance of \name{} is executed per base station (BS) as shown in Fig.~\ref{fig:slicing}, we focus our problem design and performance evaluation on a single BS characterized by a capacity $C$, which is the sum of a discrete set of available physical resource blocks (PRBs) of fixed bandwidth.
This resource availability must be divided into subsets of PRBs (i.e., slices), and our job is to dynamically assign such subsets to each network slice $i\in\mathcal{I}$. We refer to such assignment as the \emph{configuration} of slice $i$, denoted by the variable $y_i$.
Obviously, we shall guarantee $\sum_{i\in\mathcal{I}} y_i \leq C$. For the sake of clarity, we summarize all mathematical variables used throughout the paper in Table~\ref{tab:notations}.
}

We consider a time-slotted system where time is divided into \emph{decision epochs} $n = \{ 1,2,\dots,N\}$. The decision epoch duration $\epsilon$ may be decided according to the infrastructure provider policies, ranging from few seconds up to several minutes.
While the admission controller (pre-)selects a subset of slices that can co-exist without exceeding the capacity of the system \emph{in average}, the dynamic nature of the slice's load and wireless channel may cause instantaneous load surges or channel quality fading effects and hence induce a non-zero mean delay. 

%
We denote the experienced instantaneous signal-to-noise ratio (SNR) of slice $i$ (averaged out across all users of the slice) and the instantaneous aggregate traffic demand within time-slot $n$ as $\gamma_i^{(n)}$ and $\lambda_i^{(n)}$, respectively. As each tenant $i$ may show different behavior in terms of wireless channel evolution (according to $\theta_i$) and traffic demands (according to $\rho_i$), we also assume $\gamma_i^{(n)}$ and $\lambda_i^{(n)}$ are drawn from different univariate probability density function, i.e., $\gamma_i^{(n)} \sim f(x,\theta_i)$ and $\lambda_i^{(n)} \sim f(x,\rho_i)$.
Importantly, we do not assume any knowledge on such random variables; we exploit machine learning techniques to \emph{learn} the inherent channel and demand models, which allow our system to dynamically adapt the slice configurations $y_i^{(n)}$ at every decision epoch $n$ while mitigating latency constraint violations. 

Formally, the above-described problem becomes:

\vspace{2mm}\noindent \textbf{Problem}~\texttt{LATENCY-CONTROL}:
\begin{equation*}
\label{pr:latency-control}
\begin{array}{ll}
\text{minimize}   &  \lim\limits_{{N\rightarrow\infty}} \sum\limits_{n=1}^N \mathbb{E}\left [\sum\limits_{i\in\mathcal{I}} r_i^{(n)} \right ]\\
\text{subject to} & \mathbb{E}\left [ \frac{\lambda_i^{(n)}}{\zeta \left(y_i^{(n)},\gamma_i^{(n)} \right)+r_i^{(n)}}\right ]\leq \Delta_i, \quad\forall i\in\mathcal{I};\\
				  & \sum\limits_i y_i^{(n)} \geq C, \quad \forall n;\\
				  & y_i^{(n)},r_i^{(n)}\in\mathbb{Z}_+, \quad\forall i\in\mathcal{I};
\end{array}
\end{equation*}
where $\zeta(\cdot)^{(n)}$ is a mapping function that returns the number of bits that can be served using the allocated number of PRBs ($y_i^{(n)}$) and the current SNR level $\gamma_i^{(n)}$, as per~\cite[{\S7.1.7}]{TS36.213}. The traffic demand might not be satisfied within a single decision epoch incurring in packet queuing and additional delay. Therefore, in our formulation we introduce $r_i^{(n)}$ as a deficit value indicating the number of bits not served within the agreed slice latency tolerance $\Delta_i$ during the time-slot $n$ (i.e., dropped). The objective of Problem~\texttt{LATENCY-CONTROL} is hence to find a sequence of $\langle y_i^{(n)} \rangle$ configurations such that the expected total non-served traffic demand is minimized. Hereafter whenever is evident from context, we drop the superscript ${(n)}$ to reduce clutter.
\change{To address the problem, we rely on a two-layer scheduling approach commonly adopted in the network slicing context~\cite{foukas_orion,Ksentini_slicing_low_latency}. On the one side, an inter-slice scheduler is in charge of defining the PRB allocation strategy to meet the networking requirements while ensuring resource isolation among slices. On the other side, a lower layer intra-slice scheduler enforces the assignment of the pre-allocated subset of PRBs to the connected end-users. Our work mainly focuses on the higher-level inter-slice scheduler, leaving the implementation of intra-slice scheduling strategies open to address tenant-specific requirements.
}


\begin{figure}[t!]
      \centering
      \includegraphics[trim =  3.5cm 5cm 4cm 2.5cm,, clip, width=0.9\linewidth ]{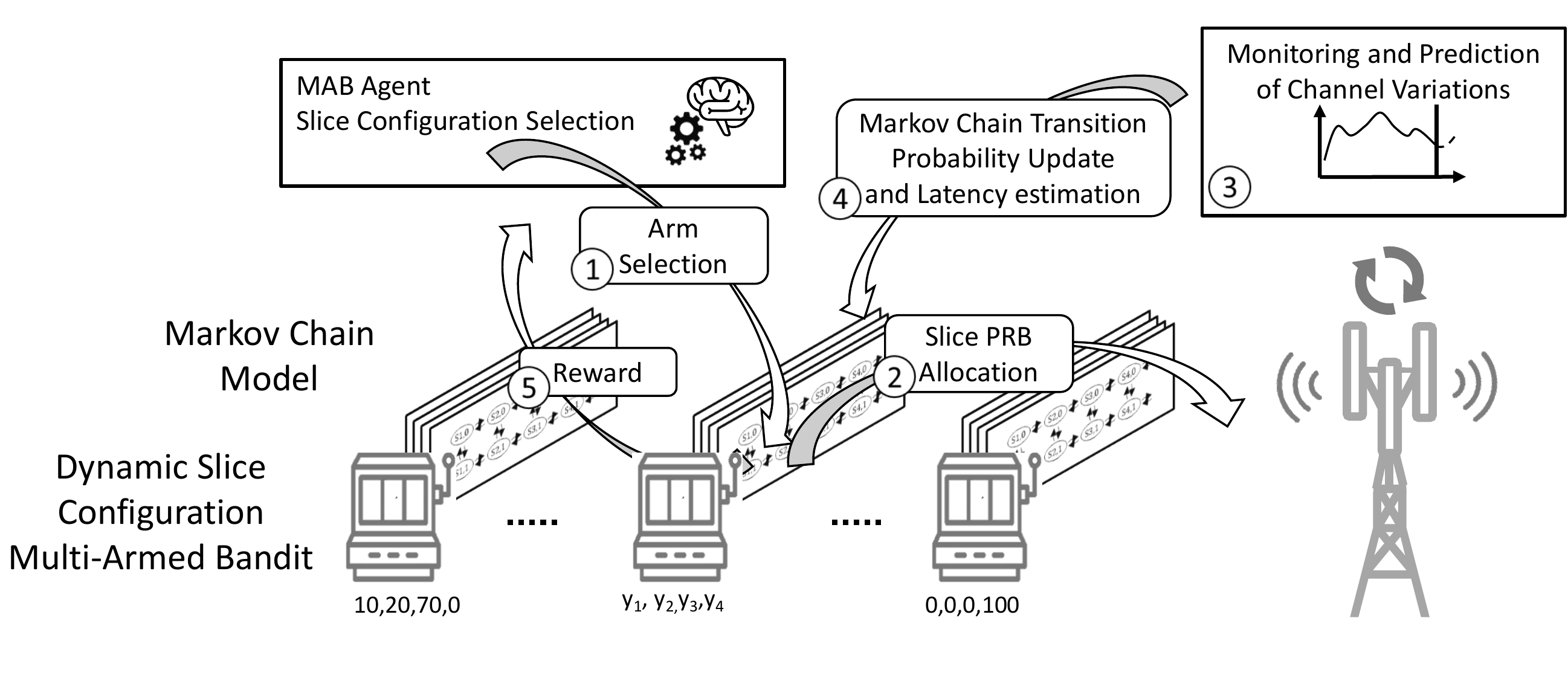}
      \vspace{-9mm}
      \caption{\small \change{Workflow illustration.}}
      \label{fig:building_blocks}
\end{figure}

\subsection{Working flow}

For a given slot $n$, problem~\texttt{LATENCY-CONTROL} can be easily linearized\footnote{Function $\zeta(\cdot)$ can be easily approximated with a linear function by applying piece-wise linearization.} and solved with standard optimization tools. However, this approach may exhibit sub-optimal behavior in future epochs if the statistical distributions of $f(x,\theta_i)$ and $f(x,\rho_i)$ are not stationary. Hence, we propose a novel two-fold approach that: $i$) models channel and traffic demand variations based on previous observations, and $ii$) iteratively applies slice settings towards the goal of honouring SLAs.

Fig.~\ref{fig:building_blocks} depicts the building blocks of our solution. \name{} relies on the concept of Markov Decision Process (MDP) as described in Section~\ref{sect:mdp} to decide which configuration $y_i$ should be enforced to all active slices $i$, adapting its choice at every epoch $n$ according to the observed \emph{reward} function that measures the incurred latency. In turn, this information is asymptotically calculated within Discrete-Time Markov Chain (DTMC) model described in Section~\ref{sect:MC}. The transition probabilities of such DTMC are updated according to previous observations in the \emph{Monitoring and Prediction of Channel Variations} module, described in Section~\ref{sect:monitor}.


\section{DTMC Model}
\label{sect:MC}
\begin{figure}[t!]
      \centering
      \includegraphics[trim = 0cm 5cm 6cm 4cm, clip, width=0.9\columnwidth ]{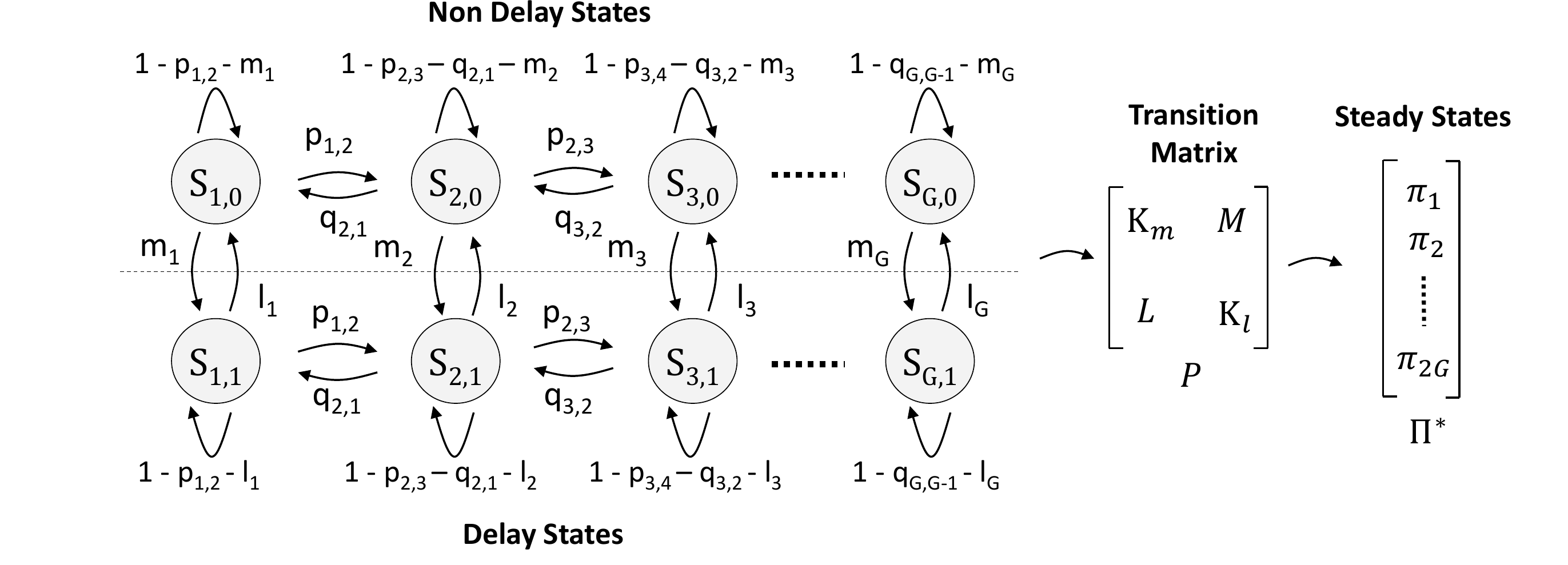}
      \vspace{-3mm}
      \caption{\change{ \small Radio channel variations as Markov chain.}}
      \vspace{-3mm}
      \label{fig:markov}
\end{figure}

Hereafter, we analyze the system dynamics through a Markov Chain-based (MC) model that computes expected channel conditions and violations on latency tolerance. It should be noted that channel variations and traffic demands are independently obtained according to each slice, thus each DTMC may be treated individually without the need to setup a Markov chain accounting for the overall system configuration. Such global DTMC could anyway be easily obtained as linear combination of the individual DTMCs. For the sake of tractability, we consider a single (virtual) user $\bar{u}_i$ with an aggregate traffic demand resulting from the set of users $u_i\in\mathcal{U}_i$ belonging to slice $i$.\footnote{This assumption can be readily relaxed by considering the convolution of single cumulative distribution functions of every user channel and demand variation~\cite{Antonic:2017:MAI:3093742.3093928}.} We also assume a finite number of channel quality levels $G$, which may bound each instantaneous user channel quality $\gamma_i$, as depicted in Fig.~\ref{fig:markov}. This is a system design choice and allows operators to trade off high accuracy for convergence speed, by ranging from a fine-grained scale (large $G$), e.g. by letting each channel quality level be equal to the modulation and coding scheme (MCSs) as defined in the 3GPP standard document~\cite{TS36.213}, to a coarse-grained scale that may capture the channel variation behaviors with limited accuracy, as detailed in Section~\ref{sect:monitor}.

\change{Let us consider a discrete-time stochastic process $X_t$\footnote{The time scale $t$ of DTMC state switch is much shorter than the decision epoch $n$ used in the MDP described in Section~\ref{sect:mdp}.} that takes values from a finite and discrete state space, which is denoted by $\mathcal{S} = \{S_{0,0}, \dots, S_{g,d}, \dots, S_{G,1} \mid 0\le g \le G, d\in\{0,1\}\}$}.\footnote{Each DTMC is defined within a state space $\mathcal{S}^i$. We remove the index $i$ to limit the clutter, as the analysis can be easily extended to any other slice $i$.} In particular, a realization of $X_t$ when visiting state $S_{g,d}$ represents virtual user $\bar{u}_i$ experiencing channel level $g\in\mathcal{G}$ with an associated delay exceeding the one specified by the slice SLA ($d=1$) or otherwise ($d=0$). 
When considering wireless channel conditions as Rayleigh distributed, it is common practice to model the variations as a sequential visiting of consecutive states, as the channel does not vary faster than the Markov chain time-slot~\cite{Munos1999}. Hence, we define the probability to improve the user channel condition from level $g$ to level $g+1$ as $p_{g,g+1}$ whereas the probability to get a bad channel from level $g$ to level $g-1$ as $q_{g,g-1}$. \change{As shown in recent works like~\cite{Verticale_Interference,Oro_Infocom_2019 }, accurate scheduling strategies might mitigate the interference effects coming from multiple base stations serving the same sets of slices thus improving the overall channel conditions. However, such schemes introduce additional complexity and synchronization overhead, which hardly fit with our view of a lightweight base station oriented solution.}
Last, given the available physical resource blocks assigned to a particular slice $y_i$, the channel quality level $g$ and the overall traffic demand within the time-slot, we model the probability to incur in delay constraint violation as $m_g$ and the probability to keep the access delay within the agreed bound as $l_g$. This process can be formulated as a two-dimensional DTMC $M := \left (\mathcal{S}, P\right)$, where $P$ denotes the following transition probability:
\begin{equation}\label{eq:trans_prob} P = \left| \begin{array}{ccc}
K_m & M \\
L & K_l \end{array} \right|,
\text{where}~K_{x=\{m,l\}} = \{k_{ij}^{(x)}\}
\end{equation}
\[\text{with}~k_{ij}^{(x)} =  
\begin{cases}
\vspace{-2mm}1-p_{i,i+1}-q_{i,i-1}-x_i  & \text{if}~i=j, \\
\vspace{-2mm}q_{i,j}  & \text{if}~i=j+1, \\
\vspace{-2mm}p_{i,j}  & \text{if}~i=j-1, \\
\vspace{-2mm}0  & \text{otherwise};
\end{cases}
\quad\text{and}~M = \{m_{i}\}_{\text{diag}}, L = \{l_{i}\}_{\text{diag}}.
\] 
Note that we assume $p_{G,G+1} = q_{1,0} = 0$ and each square block $K_{x=\{m,l\}}, M \text{ and } L$ with $\left [G \times G \right ]$ size so that the square matrix $P$ has dimension $\left[2G \times 2G\right ]$.
%
Without loss of generality, we assume that such transition probabilities do not depend on the particular time-slot we are evaluating. Thus, we define our DTMC as a time-homogeneous MC where the process $X_t$ evolves based on $\textbf{$\Pi$}(t) = \textbf{$\Pi$}(0) \textbf{$P$}^t$ where the row vectors \textbf{$\Pi$}$(t)$ and \textbf{$\Pi$}$(0)$ represent the first order state probability distribution at time $n$ and $0$, respectively.
In order to evaluate the long-term behavior of our system, we need to calculate the steady-state probability $\Pi^* = \{\pi_s^*\}$ of being in each of the defined states.
It yields that
\begin{equation}
 \textbf{$\Pi^{*}$} = \lim_{n \to \infty} \textbf{$\Pi$}(t) = \textbf{$\Pi$}(0) \lim_{n \to \infty} \textbf{$P$}^t = \textbf{$\Pi$}(0) \textbf{$P^{*}$}.
 \label{eq:limit}
\end{equation}

The above-described Markov chain is irreducible, as each state may reach through available paths any other state. Therefore, by stochastic theory, if a Markov chain is irreducible and non-periodic, the steady-state probability distribution \textbf{$\Pi^{*}$} always exists, is unique and is independent from the initial conditions. 


Recalling the total probability theorem and using Eq.~\eqref{eq:trans_prob}, we calculate the steady-state probability distribution as the solution of the following equations
\begin{equation}
\label{eq:steady_states}
\begin{cases}
    (P^T-\vv{\mathbb{1}}_{\text{diag}})\,\Pi^* & = 0  \\
    \vspace{-3mm} \vv{\mathbb{1}}\,\Pi^* & = 1
\end{cases}
\end{equation}
where $\vv{\mathbb{1}}_{\text{diag}}$ is the identity matrix.

\section{\mbox{DTMC Monitoring and Prediction}}
\label{sect:monitor}

The asymptotic behavior of a Markov chain depends on the transition probability matrix $P$, which in turn depends on the stochastic processes of the slice traffic demands and wireless channel variations. While several models have been already defined in the literature to derive such probabilities~\cite{FSMC95}, the latency control objective and the need of an accurate estimation exacerbate the problem and render model-fitting approaches impractical. This brings additional complexity and delay the convergence process to the optimal solution.

We apply the concept of \emph{unsupervised learning} to estimate the transition probabilities based on previous observations. In particular, we rely on the well-known theory of \emph{probabilistic latent variable}~\cite{latent_prob}. Let us consider $w\in\mathcal{W}$ as the stochastic latent variable denoting the current channel quality level. Formally, we redefine the transition probability of the above-described DTMC as
\begin{equation}
\rho_{a,b}^g = Pr(X_t = S_{g,b} \mid X_{t-1} = S_{g,a}, g=w)
\end{equation} 
that is the probability to move from state $S_{g,a}$ to $S_{g,b}$ when the channel level is exactly $g=w$. To easily understand this, note that $\rho_{0,1}^g = m_g$, $\rho_{1,0}^g = l_g$ whereas $\rho_{0,0}^g$ and $\rho_{1,1}^g$ are the probabilities to stay within the same state $S_{g,0}$ and $S_{g,1}$, respectively. We use an expectation maximization technique to estimate such probabilities. To this aim, we enumerate the transitions between $a$ and $b$ upon $g$ in $h^g_{a,b}$ based on the number of times $X_t$ switches to another state (or stays within the same state) between $t$ and $t+1$.
We then derive the \emph{a posteriori} probability as follows
\begin{align}
&Pr(g=w\mid X_t=S_{g,b},X_{t-1} = S_{g,a}) = \\ 
&\frac{Pr(X_t=S_{g,b}\mid X_{t-1} = S_{g,a},g=w)Pr(g=w)}{\sum\limits_{z\in\mathcal{W}} Pr(X_t=S_{g,b}\mid X_{t-1} = S_{g,a},g=z)Pr(g=z)},\nonumber
\end{align}
and the likelihood probability as the following

\begin{align}
&Pr(X_t=S_{g,b}\mid X_{t-1}=S_{g,a},g=w) =\\ &\frac{\sum\limits_{g\in\mathcal{G}}h^g_{a,b}Pr(g=w\mid X_t=S_{g,b},X_{t-1}=S_{g,a})}{\sum\limits_{\{\alpha,\beta\}\in \{0,1\}^2} \sum\limits_{g\in\mathcal{G}}h^g_{\alpha,\beta}Pr(g=w\!\!\mid\!\! X_t=S_{g,\beta},X_{t-1}=S_{g,\alpha})}\nonumber
\end{align}

and

\begin{align}
&Pr(g=w) = \\
&\frac{\sum\limits_{\{\alpha,\beta\}\in \{0,1\}^2} h^g_{\alpha,\beta}Pr(g=w\mid X_t=S_{g,\beta},X_{t-1}=S_{g,\alpha})}{\sum\limits_{\{\alpha,\beta\}\in \{0,1\}^2} h^g_{\alpha,\beta}}\nonumber
\end{align}
The above system of equations can be solved using an iterative method that yields $\rho_{a,b}^g$. Finally, we calculate the weight of each latent variable based on a given set of previous observations as per the following equation
\begin{align}
\label{eq:weights}
    &\omega(w\mid \hat{\mathcal{S}}_i) = \frac{\sum_{\{\alpha,\beta\}\in\hat{\mathcal{S}}_i} \rho_{\alpha,\beta}^w}{\sum_{g\in\mathcal{W}}\sum_{\{\alpha,\beta\}\in\hat{\mathcal{S}}_i} \rho_{\alpha,\beta}^g},
\end{align}
where $\hat{\mathcal{S}}_i$ denotes the history of transitions (or lack thereof) across $X_t$ among different states belonging to level $0$ or $1$ in the DTMC depicted in Fig.~\ref{fig:markov}. We can generalize the probability to move from a state wherein the latency is under control $S_{g,0}$ to a state incurring unexpected latency $S_{g,1}$, i.e., exceeding the threshold defined in the slice SLA, using the following expression
\begin{equation}
\label{eq:trans_prob_inf}
    \rho_{a,b}\! =\! Pr(X_{t+1}\! =\! S_b| X_{t} = S_a,\hat{\mathcal{S}}_i) =\!\!\! \sum_{w\in\mathcal{W}}\!\omega(w\!\mid\! \hat{\mathcal{S}}_i) \rho_{a,b}^w.
\end{equation}
In the next section, we design a control-theory process by means of a Markov Decision Process (MDP) that optimally selects the best slice configuration $y_i$ based on the probability to exceed the access latency constrained by the slice SLA. 

\section{Markov Decision Process}
\label{sect:mdp}
We model the decision problem as a Markov Decision Process (MDP) defined by the set of states $\Sigma = \{\sigma\}$, the set of actions $\Phi=\{\phi\}$, the transition function $T(\sigma,\phi,\sigma')$, and the reward function $R(\sigma,\phi)$.
The set of states accounts for all the radio resource splitting options among different tenants, namely \emph{slicing configuration} $c_\sigma=\{y_1,y_2,\dots,y_i,\dots,y_I\}$ expressed in terms of PRBs, where $\sum_{i\in\mathcal{I}}y_i = C$, i.e., the overall capacity is exactly split between running slices. We assume that each slicing configuration is issued at every decision epoch $n$.
The transition function characterizes the dynamics of the system from state $\sigma$ to state $\sigma'$ through action $\phi$. Analytically, $P(\sigma' \mid \sigma,\phi)$ is the probability to visit state $\sigma'$ given the previous visited state $\sigma$ and the action $\phi$.
Finally, the function $R(\sigma,\phi)$ measures the reward associated to the transition from the current state $\sigma$ performing action $\phi$.
We shall consider an MDP with an infinite time horizon. Future rewards will be discounted by a factor $0<\chi<1$ to ensure the total reward obtained is finite.

When dealing with MDPs is common practice to define a ``policy'' for the decision agent, namely a function $\mathcal{P}^{(n)}: \Sigma^{(n)} \to \Phi^{(n)}$ that specifies which action $\phi$ to perform at time $n$ when in state $\sigma$. As soon as the Markov decision process is combined with a defined policy, this automatically fixes the next action for each state so that the resulting combination exactly behaves similarly to a Markov chain.
The final aim of the decision agent is to find the policy that maximizes the expected total reward, or, equivalently, to discover the policy $\mathcal{P}^*$ that maximizes the value function. 


\subsection{Reward Definition}
\label{sect:reward_def}

Each state (or slicing configuration) is associated with a reward value that influences the agent during the decision process. The rationale behind is that we need to bind the action reward to the probability of exceeding the latency constraints defined in the slice SLA. In the following, we introduce the reward function used in our experiments with a detailed overview of its behavior.

Given a slicing configuration $c_\sigma=\{y_i \mid i\in\mathcal{I} \}$, we can analytically build a Discrete-Time Markov Chain, as described in Section~\ref{sect:MC}. If the associated transition probability matrix $P$ is perfectly known, we can also derive the steady-state probabilities $\Pi^*=\{\pi_s^*\}$ to be within any single state using Eq.~\eqref{eq:steady_states}. Thus, we can compute the probability to have the access latency of our system under control. This can be used to formulate the instantaneous reward value
\begin{equation} \label{eq:reward}
    R(\sigma^{(n)},\phi^{(n)}) = \left(\sum_{s\in\mathcal{S}_g,0} \pi^*_s\right)^\eta
\end{equation}
where $s$ is the index of all states $S_{g,0}, \forall g\in\mathcal{G}$ such that the slice latency is under control, whereas $\eta\in [0,1]$ is an adjustable value decided by the infrastructure provider to provide action fairness in the reward function when $\eta$ tends to $0$, or maximum likelihood of keeping latency under control when $\eta$ tends to $1$. Then our objective is to maximize the expected aggregate reward obtained as $\lim\limits_{{N\rightarrow\infty}} \sum\limits_{n=1}^N \mathbb{E}\left[\chi^n R\left(\sigma^{(n)},\phi^{(n)}\right) \right]$. However, given the fully-connected structure of our Markov Decision Process, i.e., all states are reachable from any MDP state, our objective is equivalent to maximize the instantaneous reward given by~\eqref{eq:reward} at each decision epoch $n$.

Nonetheless, the assumption of perfect knowledge on the transition probability matrix $P$ might be not realistic. Therefore, we need to rely on the transition probabilities $\rho_{a,b}$ inferred based on the previous observations, as explained in Section~\ref{sect:monitor}, Eq.~\eqref{eq:trans_prob_inf}. The larger the set of observations, the higher the accuracy of our probability estimation and the higher the reward attained to the instantaneous best action taken by the MDP.

\subsection{Complexity analysis}

\change{Once we have fully characterized our proposed MDP, we can solve it by using dynamic programming solutions such as Value Iteration~\cite{valueiteration}. These approaches require exploring the entire state space of the MDP (several times) and the associated rewards. 
Let us consider a scenario with $I$ online slices running in our system. Assume that each slice configuration $y_i$ can take values from integer multiples of a minimum PRB chunk size $\Theta$ and that the slicing configuration must be consistent, i.e., $\sum_{i\in\mathcal{I}}y_i = C$. Then, we can calculate the overall number of states equal to $\frac{( \frac{C}{\Theta}  +I-1)!}{(I-1)! \frac{C}{\Theta}!}$.
This poor state scalability, as well known as \emph{the curse of dimensionality}, compromises the feasibility of MDP models under practical conditions. However, MDPs provide insights regarding the structure of the problem itself and are very helpful to design ausiliary solutions, such as Multi-Armed Bandit (MAB) models, which are better suited for functional deployments. Therefore, in the next section we rely on a novel MAB design that exploits information from the underlying MDP to expedite the learning process while attaining near-optimal results.
}

\subsection{Multi-armed Bandit problem}
\label{sect:mabmodel}

The online decision-making problem has been addressed in the past with several mathematical tools~\cite{survery_mab}. The limited information about real-time channel quality and effective traffic demand forces the operator to choose, like a gambler facing diverse options to play, the number of radio resources to assign to each running slice. This automatically falls in the fundamental \emph{exploration-vs-exploitation} dilemma: the gambler needs to carefully balance the exploitation operations on known slicing configurations that provided the best revenues in the past against the exploration of new slicing configuration that might eventually produce higher revenues.

This class of decision process can be formulated as a Multi-Armed Bandit (MAB) problem, which emulates the action of selecting the best (single) bandit (or slot machine) that may return the best payoff. Each slot machine returns unpredictable revenues out of fixed statistical distribution, not known \emph{a priori}, that is iteratively inferred by previous observations. This matches well the randomness of the channel quality and the traffic demand we aim to capture whereas each bandit can be mapped onto a state of the MDP, i.e., a specific slicing configuration. The final objective of such a problem is to maximize the overall gain after a finite number of rounds. This class of problems is usually assessed by a defined metric called \emph{regret} $\Omega$, which is defined as the difference between the reward that can be gained by an optimal oracle, i.e., using an optimal policy that knows the reward distributions \emph{a priori}, and the expected reward of the myopic online policy.

Reusing notation from our MDP model, let us define each arm $\sigma \in \Sigma$ as a different slicing configuration $c_\sigma = \{y_i \mid i \in\mathcal{I}\}$. Once selected, each arm provides an instantaneous reward $R(\sigma)$ defined as the following
\begin{equation}\label{eq:classic_mab_reward}
    R(\sigma) = \sum_{i\in\mathcal{I}}\left(\zeta(y_i,\gamma_i)-\frac{\lambda_i}{\Delta_i}\right)
\end{equation}
where the slicing configuration is $y_i\in c_\sigma$, $\zeta(\cdot)$ computes the number of bits that can be served using $y_i$ configuration and given the current channel quality $\gamma_i$, and $\lambda_i$ is the slice traffic demand, as described in Section~\ref{sect:prob_def}.

While using such reward function requires low overhead, as it only needs to calculate the incurred latency after selecting a slicing configuration, it only converges to a near-optimal solution after exploring several configurations, which results in overly long training periods (as shown in Section~\ref{sect:perf_eval}). This is an inherent issue with classic MAB methods, which are \emph{blind} to the underlying system structure. Conversely, in this paper we resort to a novel model-assisted approach that exploits the system model of Section~\ref{sect:reward_def} to guide the exploration/exploitation process with (abstract) system information. In this way, as opposed to using the traditional reward model of Eq.~\eqref{eq:classic_mab_reward}, we define our bandit's reward as the expectation of access latency exceeding slice SLA defined in Eq.~\eqref{eq:reward}. This has a two-fold advantage: $i$) during the initial training period, the DTMC associated to each state of the MDP is updated (and enhanced) with more accurate values of the transition probabilities: this helps to find steady-state probabilities (and in turn an updated reward per slicing configuration) that reflect the real behavior of our system as time goes on; and $ii$) the slicing configuration selection accounts directly for stochastic behaviors of both channel quality and traffic demand, while reducing the state space to those that may benefit the entire system. Many algorithms have been proposed to optimally solve the MAB while learning from previous observations~\cite{Vermorel2005}. One of the main issues is that collecting rewards on a short-time basis may negatively impact on the decision of the best bandit. Thus, we rely on a modified version of the so-called \emph{Upper Confidence Bound} (UCB) algorithm devised by~\cite{AuerMab2002} that overcomes this issue by measuring not only the rewards collected up to the current time interval, but also the \emph{confidence} in the reward distribution estimations by keeping track of how many times each bandit has been selected $z_{\sigma,n}$. The pseudo-code is listed in Algorithm~\ref{alg:ucb}.

\begin{algorithm}[t!]
\caption{\change{\small{\textbf{\name{}}}}}
\vspace{-6mm}
\label{alg:ucb}
\algsetup{indent=2em, linenodelimiter=.}
\begin{footnotesize}
\begin{multicols}{2}
\begin{algorithmic}[1]
\STATE {\bf Input: }$ \Sigma, N, \Psi =\{\psi(\sigma)\}, \change{\mathcal{I}, \omega , \epsilon, \mathcal{S}} $
\STATE {\bf Initialization: }$  z_{\sigma},\hat{\rho}_{\sigma} = 0,\,\,\forall \sigma$
\STATE {\bf Procedure:}
\FORALL {$n \in N$}
\IF {$n = 0$} \label{line}
    \FORALL {$\sigma \in \Sigma$}
    \STATE GET reward: $\hat{\rho}_{\sigma}=  R_{\sigma}^{(n)}$
    \STATE $z_{\sigma} = z_{\sigma} + 1$
    \ENDFOR
\ELSE
	\STATE $\sigma^* = \argmax\limits_{\sigma\in\Sigma}\hat{\rho}_{\sigma} + \psi(\sigma) \sqrt{\frac{2\log{\sum_k z_{k}} }{z_{\sigma}}}$
	\STATE UPDATE $\hat{\rho}_{\sigma^*} \leftarrow R_{\sigma^*}^{(n)}$
	\STATE $z_{\sigma} = z_{\sigma} + 1$
\ENDIF
\change{
\FORALL {TTIs $\in \epsilon$}
    \FORALL {$i \in \mathcal{I} $}
        \STATE UPDATE $\omega(w\mid\hat{\mathcal{S}}_i)  \leftarrow \mathcal{S}_i$
    \ENDFOR
\ENDFOR
\STATE UPDATE $\psi(\sigma^*)\leftarrow\omega(\cdot)$
}  


\ENDFOR
\STATE {\bf End Procedure}
\end{algorithmic}
\end{multicols}
\end{footnotesize}
\end{algorithm}

 Initially, we explore all bandits, i.e., slicing configuration $\sigma\in\Sigma$, to get a consistent reward (line 2-6). Then we select the best configuration that maximizes the empirical distribution $\hat\rho_\sigma$ accounting for a confidence value. This confidence value depends on the number of times we have explored that particular configuration as well as the accuracy of the transition probabilities we calculate for the associated DTMC. Note that this is different to traditional UCB algorithms. Specifically, we define a Markov accuracy value $\psi(\sigma) = (\frac{(\sum_w \omega(w\mid\hat{\mathcal{S}}_i))^2}{W\sum_{w}\omega(w\mid\hat{\mathcal{S}}_i)^2})$, where $W$ represents the cardinality of the set $\mathcal{W}$. Note that $\psi(\sigma)$ depends on the weights $\omega(\cdot)$ obtained through the performed observations $\hat{\mathcal{S}}_i$, as reported in Eq.~\eqref{eq:weights}. Interestingly, $\psi(\sigma)\in(0,1]$, i.e., when the DTMC has no relevant observations to build its transition probabilities this function returns $\psi(\sigma) =1$ whereas, when a relevant number of observations allow to determine accurate transition probabilities, its value tends to $0$. \change{The value of $\psi(\sigma)$ is updated at the end of every decision interval (line $20$) after monitoring the effects of the last decision on the Markov Latent variable distribution (lines $15-19$). }





    
    
    
    
    
    

\subsection{Regret analysis}
Here, we mathematically calculate the bounds of our solution, \name, for multi-armed bandit problems. Let us consider a player selecting an arm $\sigma\in\Sigma$ every decision epoch $n$. Every time arm $\sigma$ is pulled down, it returns a reward $R_{\sigma}^{(n)}$ drawn from an unknown distribution with mean $\bar{\rho}_\sigma$ and empirical mean value calculated until time $n$ as $\hat{\rho}_{\sigma}^{(n)}=\frac{\sum_{s=1}^n R_{\sigma}^{(s)}}{n}$. We denote $\sigma^*$ as the arm providing the maximum average reward such that $\bar{\rho}_{\sigma^*}>\bar{\rho}_\sigma, \forall\sigma\neq\sigma^*$.
If the arm selection is performed using \name, it yields that the regret is obtained as
\begin{align} \label{eq:regret}
\Omega_N^{\text{\name{}}}(\Sigma) & = N\bar{\rho}_{\sigma^*}-\mathbb{E}[\sum\limits_{n=1}^NR_{\sigma}^{(n)}\mid \sigma\in\mathcal{P}^{\text{\name{}}}] \nonumber\\
				  & = N\bar{\rho}_{\sigma^*}-\sum\limits_{\sigma=1}^{\Sigma} \bar{\rho}_{\sigma} \mathbb{E}[z_{\sigma}^{(n)}];
\end{align}
where $\mathcal{P}^{\text{\name{}}} = \{{\sigma_n}\}$ is the policy as defined in Section~\ref{sect:mdp} that consists of a set of moves that \name{} will play at time $n$ whereas $z_{\sigma,n}$ is the overall number of decision epochs arm $\sigma$ has been pulled down till time instant $n$. Now consider \name{} as a uniformly good policy, i.e., any suboptimal arm $\sigma\neq{\sigma^*}$ is chosen by our policy up to round $n$ so that $\mathbb{E}[z_{\sigma,n}] = o(n^\alpha),\forall \alpha>0$. It holds that
\begin{equation}
\lim_{\text{N}\to\infty} \sum\limits_{\sigma=1}^{\Sigma} N^{-1}\bar{\rho_\sigma} \mathbb{E}[z_{\sigma}^{(N)}] = \Sigma \bar{\rho}_{\sigma^*}. 
\end{equation}
Hence, we can express the regret lower bound as the following
\begin{equation}
\lim_{\text{N}\to\infty}\inf \frac{\Omega_N^{\text{\name{}}}(\Sigma)}{\log N} \geq \!\!\!\!\sum\limits_{\sigma:\,\, \bar{\rho}_\sigma<\bar{\rho}_{\sigma^*}} \!\frac{\bar{\rho}_{\sigma^*}-\bar{\rho}_{\sigma}}{Div(\bar{\rho}_\sigma,\bar{\rho}_{\sigma^*})}
\end{equation}
where $Div(\bar{\rho}_\sigma,\bar{\rho}_{\sigma^*})$ is the Kullback-Leibler divergence of one statistical distribution against the other and it is used to measure how one distribution might diverge from another probability distribution.

Now consider the Hoeffding's inequality for multiple i.i.d. variables $x_n$ with mean $\mu$. It yields that $Pr(|\frac{\sum_{i=1}^n x_i}{n}-\mu|\geq\delta)\leq 2e^{-2n\delta^2}$. Our algorithm \name{} applies an upper confidence interval $\delta = \sqrt{\frac{2\log\sigma_kz_k}{z_\sigma}}$. Therefore, it yields that
\begin{equation}
    Pr\left(|\hat{\rho}_{\sigma,n}-\bar{\rho}_\sigma|<\sqrt{\frac{2\log\sum_kz_k}{z_{\sigma}}}\right) \geq 1-\frac{2}{n^4}
\end{equation}
and also that
\begin{equation}
    Pr\left(\mathcal{P}^{(n+1)}=\sigma\mid z_{\sigma}^{(n)}>\frac{4\log n}{\bar{\rho}_{\sigma^*}-\bar{\rho}_{\sigma}}\right) \leq \frac{4}{n^4}.
\end{equation}
We can then derive the expectation of number of times sub-optimal arm $\sigma\neq{\sigma^*}$ is pulled down as follows
\begin{equation}
    \mathbb{E}[z_{\sigma}^{(N)}] \leq \frac{4\log N}{\bar{\rho}_{\sigma^*}-\bar{\rho}_{\sigma}}+8
\end{equation}
and the regret upper bound as the following
\begin{equation}
    \mathbb{E}\left[\Omega_N^{\text{\name{}}}(\Sigma)\right]\leq \sum\limits_{\sigma:\,\, \bar{\rho}_\sigma<\bar{\rho}_{\sigma^*}}\frac{4\log N}{\bar{\rho}_{\sigma^*}-\bar{\rho}_{\sigma}}+8\left(\bar{\rho}_{\sigma^*}-\bar{\rho}_{\sigma}\right).
\end{equation}


\section{Performance Evaluation}
\label{sect:perf_eval}





In this section, we evaluate our solution through an exhaustive simulation campaign that takes into account complexity, revenue and SLA violation metrics. 

\subsection{Simulations setup}
To assess heterogeneous slices, we simulate the network load demand of slice $i$ at each time-slot (i.e., each transmission time interval (TTI) in Long Term Evolution (LTE) systems) by extracting a random value from a Normal distribution $\mathcal{N}_i (\mu_i ,\nu_i^2 )$, where $\mu_i$ and $\nu_i$ represent the mean value and standard deviation, and let $L_i$ describe its latency constraint. Moreover, we model the SNR channel variation as another random variable drawn by a Rayleigh distribution and derive the probability distribution encompassing the whole SNR range.
For every channel instantiation, we extract the corresponding Modulation and Coding Scheme (MCS) as defined by the 3GPP standard.\footnote{We refer the reader to~\cite{TS36.213} for an exhaustive explanation of the mapping between SNR and MCS.} The MCS index $m \in \mathcal{M}$ combines one possible modulation scheme and a predefined coding rate providing a compact way to represent a simple concept: the better the radio conditions, the more bits can be transmitted per time unit, and \emph{vice versa}. 
Fixing the channel bandwidth, the expected average throughput achievable by one slice during one epoch depends on both the modulation and coding schemes used and, most importantly, on the number of PRBs reserved for the slice. In a wider timescale\,\footnote{Note that we assume a timescale larger than our epochs used in the decision-making process.}, the average capacity can be approximated as $C_i = \left( \sum_{m}^{M} \Gamma_{m} \pi_{m,i} \right) T_i y_i$ where $\Gamma_{m}$ represents the average number of bits per LTE subframe that can be transmitted using the $m$-th MCS index, $\pi_{m,i}$ is the steady-state probability distribution output of the first stage Markov chain model, $T_i$ defines the decision interval size, and $y_i$ accounts for the number of PRBs allocated to the $i$-th slice. We refer the reader to Table~\ref{tab:notations}.
\change{In the LTE radio interface, the maximum amount of PRBs is fixed to $100$ when operating at conventional bandwidth values of $20$ MHz. In order to support massive type communication and Ultra-Reliable Low-Latency Communication (URLLC) use-cases, the 5G New Radio (NR) introduces significant enhancements in the radio frame composition. Not only 5G NR will support wider channel bandwidth (up to $100$ MHz), but also introduce the support for multiple different types of subcarrier spacing. For back-compatibility reasons, even in 5G NR the time duration of radio frames and subframes are fixed to $10$ ms and $1$ ms, respectively\cite{TS38.211}. The number of slots within each subframe however would change according to the subcarrier configuration, which eventually translates in shorter PRB time duration and thus a different PRB availability depending on the selected configuration. It must be noticed that all the subcarrier spacing are defined as $\Delta f = 2^j \cdot 15$ KHz, $j=\{0,\dots,4\}$, thus leading at the definition of time-frequency grids containing an amount of PRBs which is multiple of those contained in the traditional LTE grids. In this context, we assume a simple mapping function, as the one described in~\cite{lasr}, implemented at intra-slice scheduler to homogenize the resources of potentially heterogeneous radio access technologies.}

Traffic demands are compared with the current channel availability to derive the possibilities to pass from one state to another. It must be noticed that the accuracy of the resulting steady-state distribution strictly depends on the precision of such comparison. For this reason, we constantly monitor and update the transition probabilities of the Markov chain based on the resource allocation adopted in the current decision interval. During the arm selection, if the chosen configuration does not provide enough resources to meet the latency requirements, the steady-states will be mostly distributed in the lower part of the Markov chain leading to a minor reward that, in turn, guides the MAB agent to take a different action (i.e., selecting a different arm) in the following decision round.

For benchmarking purposes, we implement two widely used MAB algorithms, namely ``legacy'' UCB and Thompson Sampling (TS)\footnote{Due to space limits, we refer the reader to the literature introducing such algorithms, e.g.~\cite{Agrawal2017}.}. On the one hand, UCB adopts a deterministic approach to deal with the exploration-vs-exploitation dilemma, but its performance generally degrades as the number of arms increases. On the other hand, Thompson sampling adopts a probabilistic approach that scales better with the number of arms, but it may provide sub-optimal results when the distribution of reward changes over time (i.e., in non-stationary scenarios). Conversely, \name{} combines the advantages of them both by adopting a probabilistic model (MDP) guiding an exploration phase derived from UCB.

\subsection{Multi-armed bandit problem behavior}
\label{sec:MAB_behaviour}
We first explore the trade-off between action space (and its granularity) and the associated reward loss. To this aim, we set up a simple experiment with $2$ slices with equal SLA requirements in a deterministic and static environment. We then apply \name{} using $3$ different action sets: $\{0, 2, 4, \dots, 100\}$, $\{0, 5, 10, \dots, 100\}$ and $\{0, 10, 20, \dots, 100\}$ PRBs (with $50$, $20$ and $10$ available configurations each), labelled ``2 PRBs``, ``5 PRBs`` and ``10 PRBs``, respectively. The results, shown in Fig.~\ref{fig:prbs} make it evident that the higher the granularity the longer the exploration phase(s): over $50$ intervals for ``2 PRBs`` whereas it takes around $10$ intervals for ``10 PRBs``. Interestingly, the loss in reward attained to the latter configuration is only $2\%$. Therefore, due to a faster convergence time at the expense of minimal reward loss, we empirically select $\Theta=10$ PRBs for our purposes.


\subsection{Slice SLA violation analysis}

\begin{figure}
\centering
\begin{minipage}{.5\textwidth}
  \centering
      \includegraphics[trim = 1.5cm 0cm 1cm 0cm , clip, width=\columnwidth]{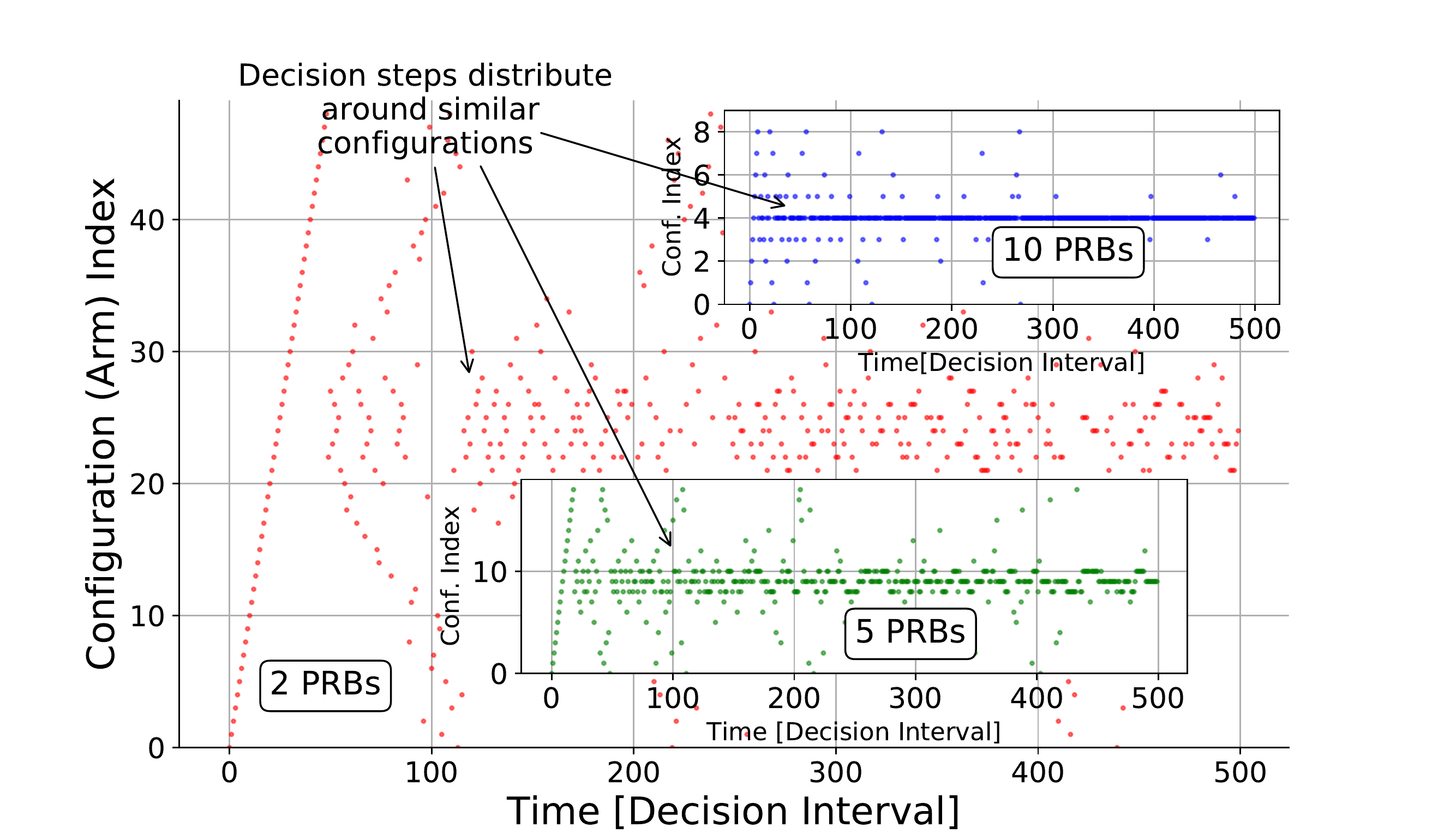}
      \vspace{-10mm}
      \caption{\small Impact of different resource allocation \\ chunk sizes.}
      \label{fig:prbs}
      \vspace{-6mm}
\end{minipage}%
\begin{minipage}{.5\textwidth}
  \centering
      \includegraphics[trim = 1cm 0cm 1cm 0cm , clip, width=\columnwidth]{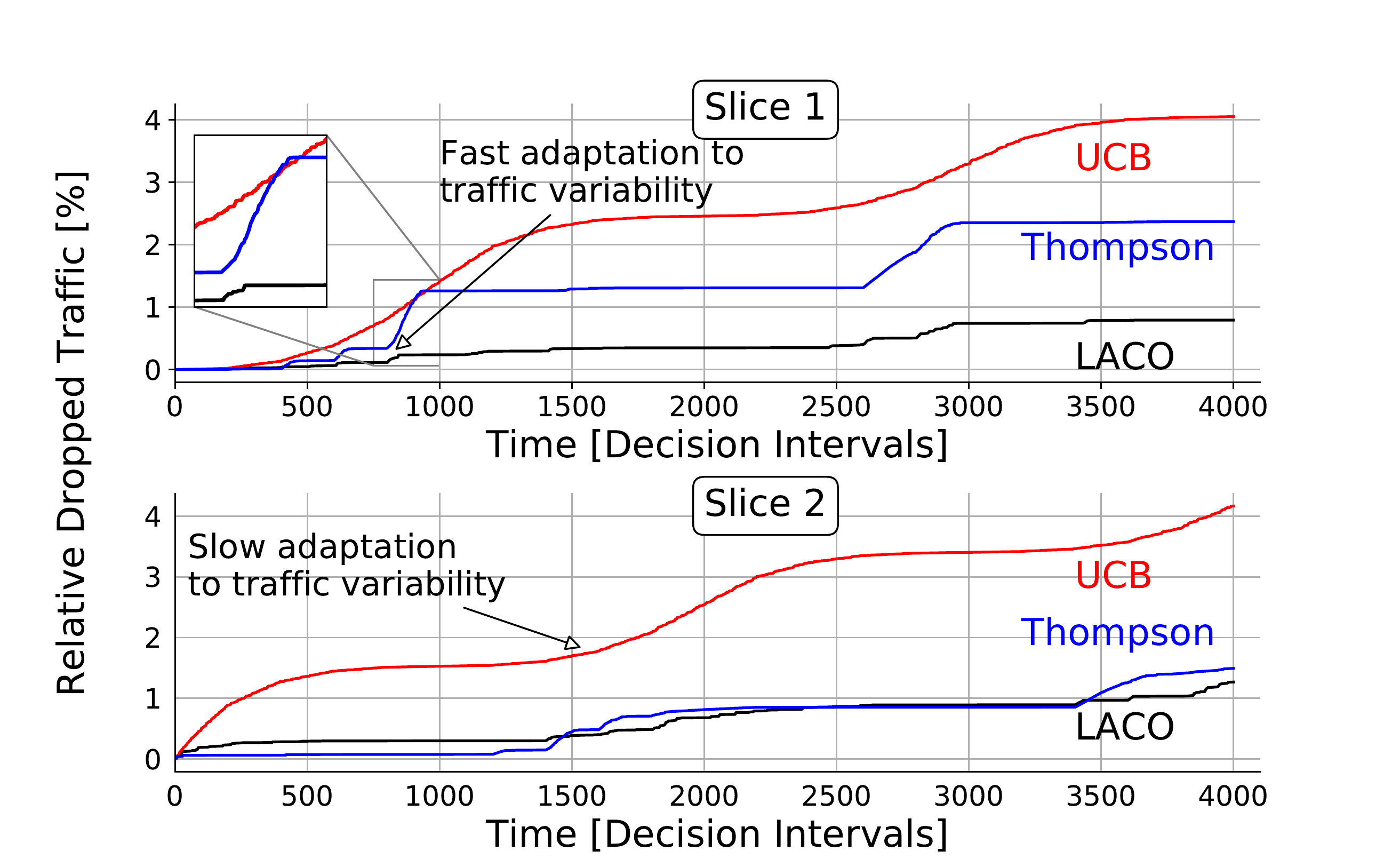}
      \vspace{-10mm}
      \caption{\small Cumulative dropped traffic due to latency constraints violations.}
      \label{fig:cumulative_dropped}
      \vspace{-6mm}
\end{minipage}
\end{figure}

We thus grant spectrum-time resources in the granularity of chunks of $1~\text{second}\times10~\text{PRBs}$.
In the first scenario, we investigate the capacity of \name{} to adapt the resource allocation at variable traffic loads. For this reason, we consider only two slices with equal requirements, i.e., ${\nu_i^2}=10$ Mb/s and $\Delta_i=20$ ms for $i={1,2}$. To assess real scenarios with non-stationary traffic patterns, we vary the mean load of each slice $i$ following a sinusoidal curve in counter-phase between $\mu_i=8$ Mb/s and $\mu_i=40$ Mb/s. This forces the resource allocation process to span across the whole configuration set when dealing with SLAs guarantees. As shown in Fig.~\ref{fig:cumulative_dropped}, the cumulative dropped traffic of each slice changes when different MAB algorithms are used. The behaviour of UCB shows high variability after few decision intervals. As soon as all the arms are selected, the agent starts learning about the statistics of the outcomes and builds a ranked list. The need for a comprehensive knowledge of all the arms leads to several "bad" choices during the exploration phase. This slows down the convergence to the optimal configuration and penalizes performance. From the obtained rewards, TS builds a bivariate probability distribution across the expected reward of each arm, extracts a random sample and chooses the arm associated to the maximum value. This approach performs well in static scenarios as TS favours exploitation of the empirical results obtained in the first attempts; but in time-varying scenario as the one we are considering, the reward distribution associated to each arm fluctuates over time rendering TS unable to adapt fast enough in highly-dynamic scenarios. In contrast, the \name{}'s model-awareness allows for quicker convergence and so it accommodates real-time traffic requirements in dynamic environments and as a result reduces the amount of data violating delay deadlines.


\begin{figure*}[t!]
    \centering
	  \begin{subfigure}[b]{.33\textwidth}
          \includegraphics[trim = 0cm 0cm 0cm 0cm,clip, width=\linewidth]{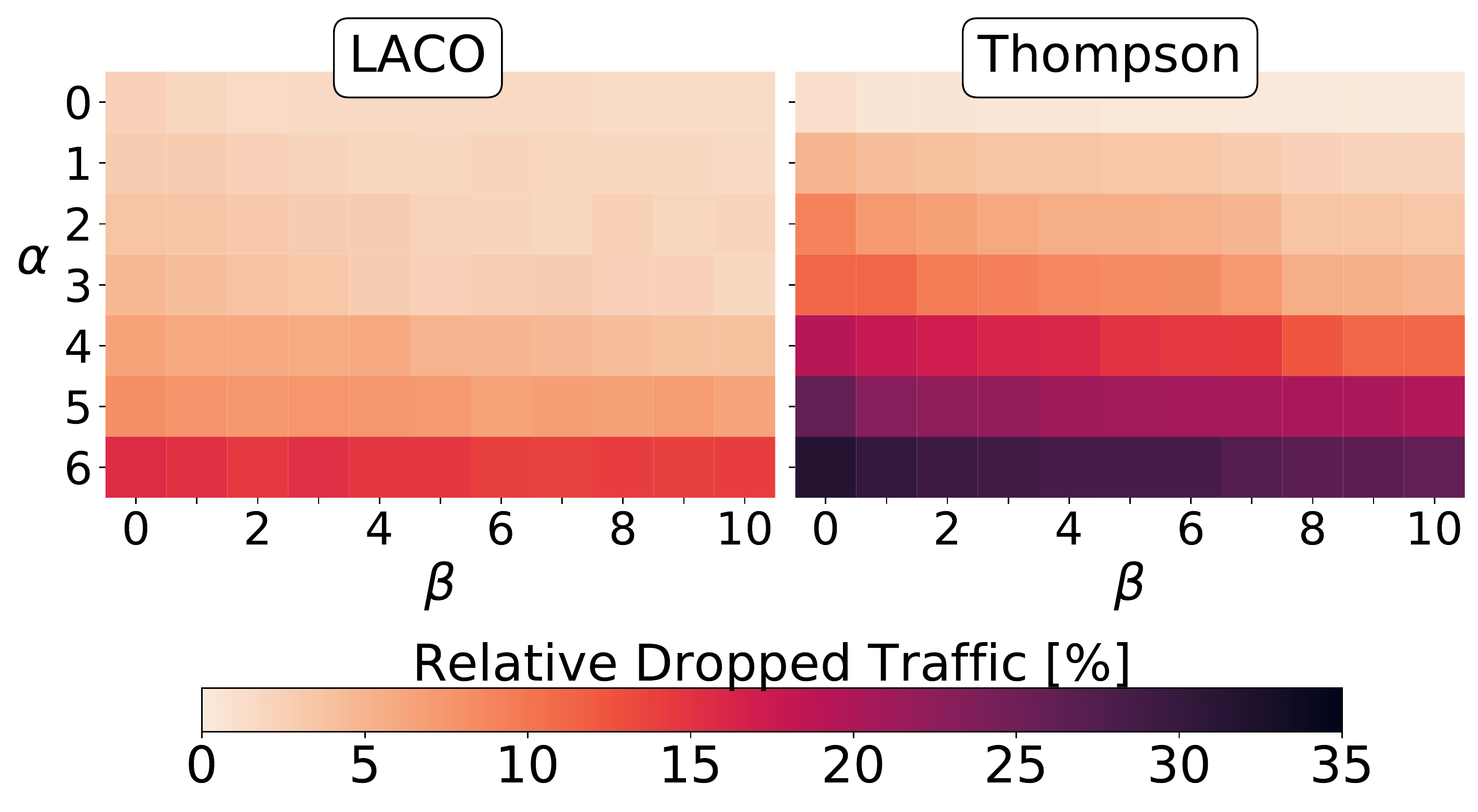}
          \caption{\small Effects of different slice reqs.}
          \label{fig:heatmap}
      \end{subfigure}%
	  \begin{subfigure}[b]{.32\textwidth}
        \includegraphics[trim = 0cm 0cm 0cm 0cm, clip,width=\linewidth]{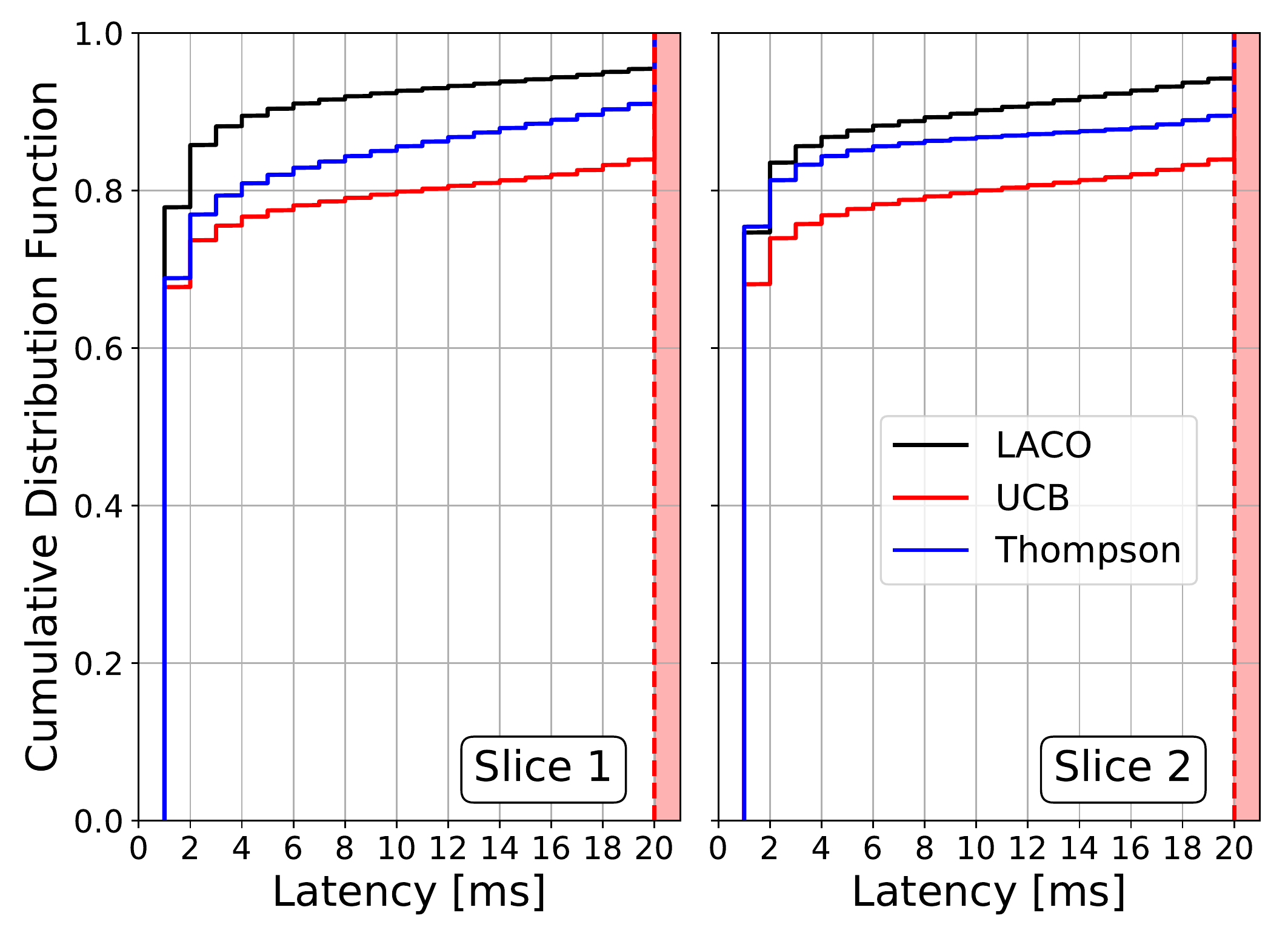}
        \caption{\small CDF of experienced latency.}
        \label{fig:cdf}
      \end{subfigure}%
	  \begin{subfigure}[b]{.32\textwidth}
        \includegraphics[trim = 0.5cm 0cm 1.5cm 0cm,clip,width=\linewidth]{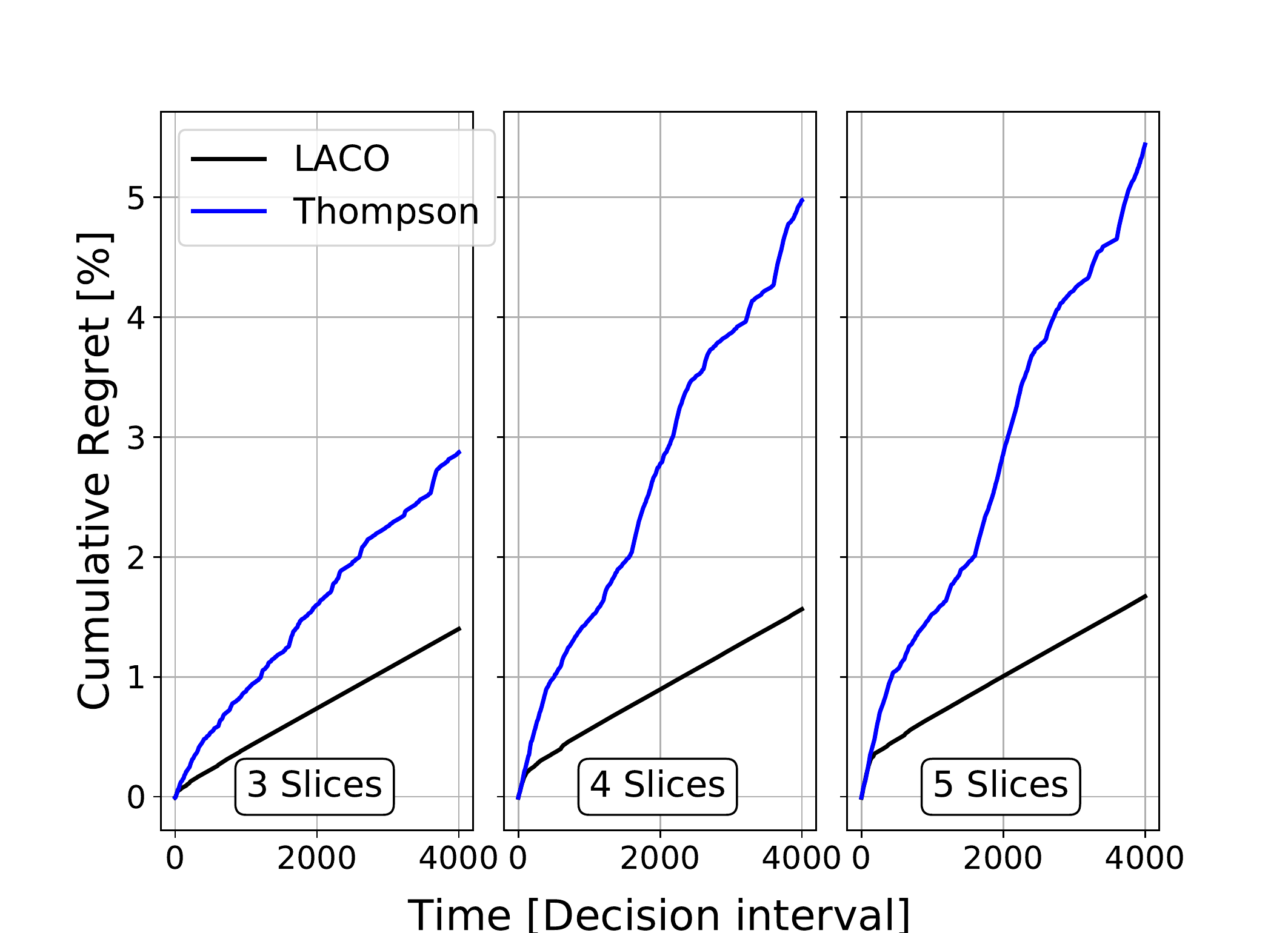}
        \caption{\small Empirical cumulative regret.}
        \label{fig:cumulative}
      \end{subfigure}%
    \vspace{-2mm}
    \caption{\small (a) Effects of different slice requirements; (b) CDF of latency experienced by served traffic; (c) Empirical cumulative regret for a variable number of slices.}
    \vspace{-4mm}
	\label{fig:Sims_results}
\end{figure*}

Obviously, heterogeneous throughput/latency requirements impact the system differently. Fig.~\ref{fig:heatmap} shows the effect of such variations on the system extending the previous scenario and considering increasing values of resource requirements as $10\cdot\alpha$ Mb/s, and $10\cdot\beta$ ms, respectively. As expected, smoother delay requirements (horizontal direction in the figure) allow to serve more traffic within the latency bounds defined by the SLA, although the impact becomes negligible after few incremental steps. This is due to long decision intervals when compared to the timescale of fast channel variations. A proper resource configuration selection allows to match the offered traffic requirements with the expected channel capacity, allowing the incoming traffic to be served within few milliseconds. As the offered traffic approaches the channel capacity boundary (vertical direction in the figure), the same task becomes more challenging and the admission and control process should consider this aspect when granting/rejecting access to new network slices.
\name~'s abilities to adapt to demand variations not only mitigates the amount of traffic violating delay requirements but also improves the distribution of data delivery delay overall. As shown by Fig.~\ref{fig:cdf}, the empirical CDF of delay for each slice in the same scenario presented above remarkably improved with a mean delay equal to $2.6$, $3.9$ and $4.9$ ms for \name{}, TS and UCB, respectively.

Finally, we implement an optimal offline policy with full knowledge of the system, i.e., an oracle policy that knows the future with the corresponding latency violations. We compare both \name{} and TS to this optimal policy for a variable number of slices. The aggregated demand is adapted to ensure we operate within the system capacity. In Fig.~\ref{fig:cumulative}, we depict the temporal evolution of the cumulative reward loss over time (regret) for both approaches. The figure illustrates how the regret increases with time much rapidly for TS, a difference that increases with the number of slices.

\subsection{ \change{Convergence time}}
\change{
The next generation of mobile networks (5G) promises to support the provisioning of high throughput and low-latency services even in highly dense scenarios~\cite{Slicing2018CoNEXT}. These capabilities are tightly bounded with the possibility to exploit higher communication frequencies together with wider spectrum bandwidth. In the 5G context, bandwidth is expected to increase up to $100$MHz, leading to additional complexity in the management of radio resources.
In order to assess \name{} performances in such scenarios, we investigate the convergence time of our solution to the optimal slice configuration in different bandwidth settings. To enable more efficient use of the spectrum resources and reduce the power consumption at UE side, 5G New Radio (NR) introduces the concept of bandwidth parts (BWP)~\cite{TS38.211}, where each BWP can be configured by different numerologies defining specific signal characteristic, e.g., in terms of subcarrier spacing. Without loss of generality, we assume all the end-users belonging to the same slice operating under similar numerology settings. Moreover, we keep the subcarrier spacing fixed to $\Delta f =15$ KHz as in legacy LTE systems. Such coarse resource allocation scheme is mandatory to support LTE devices but, it can be easily mapped to finer resource block structures as defined within the 5G domain at lower layer intra-slice schedulers~\cite{lasr}.

Fig.~\ref{fig:convergence_vs_N_slices} compares the convergence time of different MAB algorithms for an increasing number of slices and bandwidth availability over a time period of $N=1000$ decision intervals.
It should be noticed that depending on channel statistics and real-time slice requirements, \emph{multiple} resource allocation settings (namely arms) may provide \emph{optimal performance} making unfeasible a single convergence point. Thus, we opted to simulate the worst-case scenario allowing for a unique optimal resource configuration in each simulation run. In line with previous observations, we fix $\Theta = 0.1\,C$.
Despite a common initial exploration phase (highlighted in orange), from the picture it is evident how the curse of dimensionality affects the overall convergence time. This is more evident for the legacy UCB approach (depicted in red), which hardly copes with the increasing size of the action space and in some runs did not converge to a solution within the time boundary of our experiment. Focusing on \name{} performances (depicted in black), the number of decision intervals necessary to converge to the optimal resource allocation outperforms Thompson Sampling (in blue) by scaling almost linearly with the number of slices (and PRB availability) after the initial exploration phase.

\begin{figure*}[t!]
	\centering
	\begin{subfigure}[b]{.49\textwidth}
        \includegraphics[trim = 0cm 0cm 0cm 0cm, clip, width=\linewidth ]{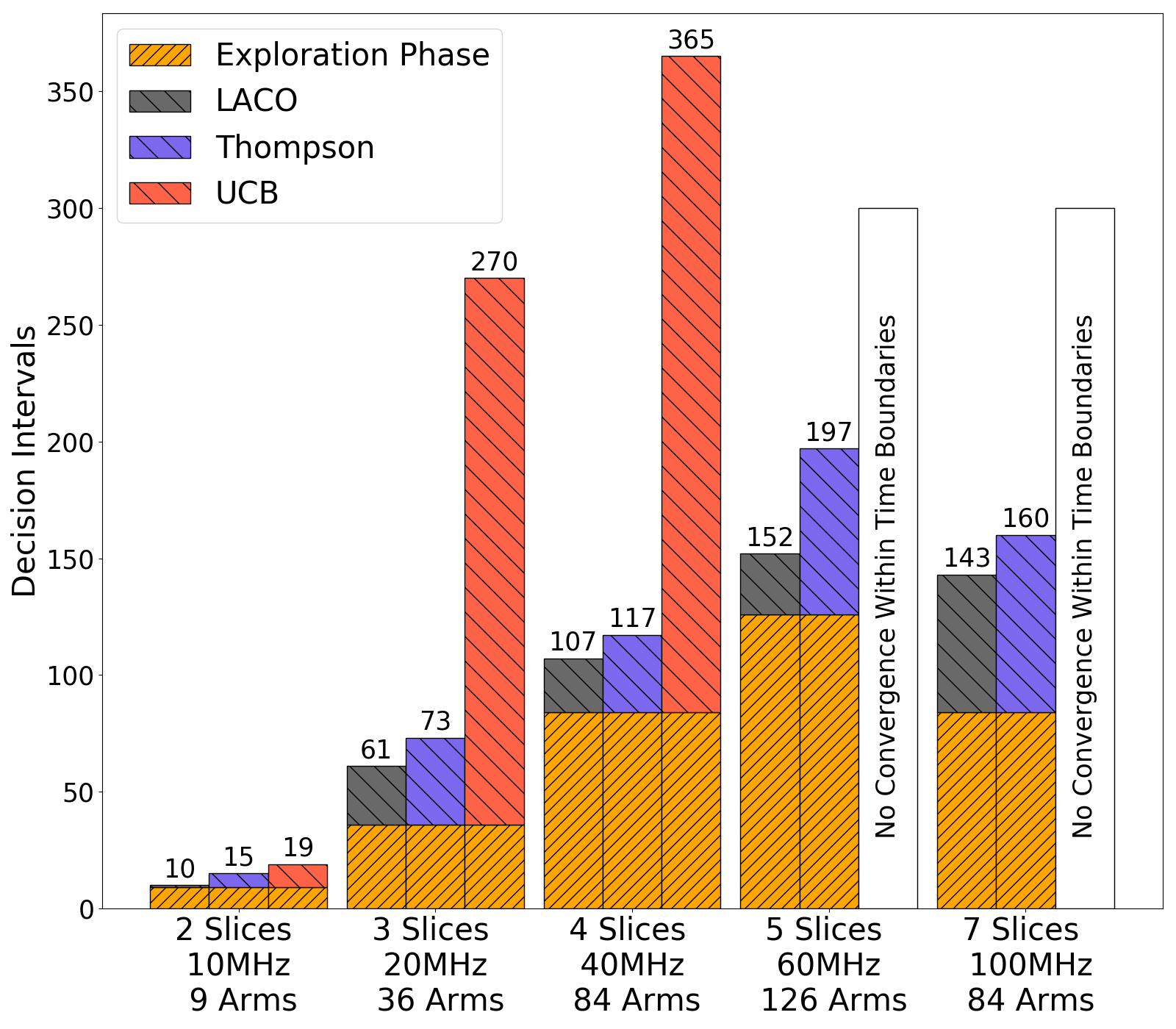}
        \caption{\small \change{Effects of different number of slices and \\bandwidth (PRB) availability.}}
        \label{fig:convergence_vs_N_slices}
    \end{subfigure}%
    \begin{subfigure}[b]{.49\textwidth}
        \includegraphics[trim = 0cm 0cm 0cm 0cm, clip, width=\linewidth ]{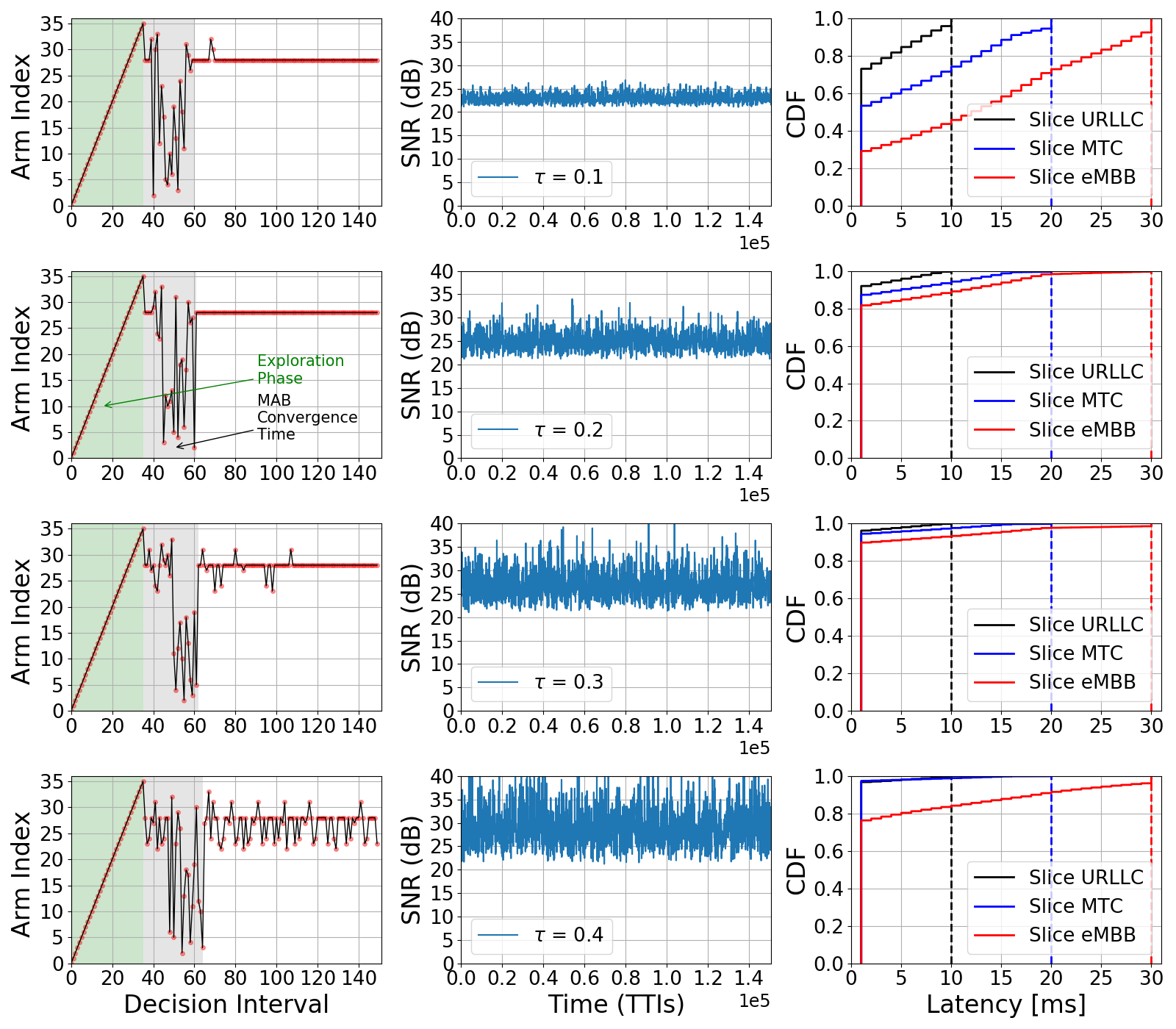}
        \caption{\small \change{Effects of increasing variability in the channel conditions (SNR).}}
        \label{fig:convergence_vs_SNR}
    \end{subfigure}%
    \caption{\small \change{Sensitivity analysis of bandwidth availability and SNR variability on the convergence time to the optimal slice resource allocation.}}
        \vspace{-4mm}
\end{figure*}%

Convergence to the optimal slice configuration also shows its dependency on the radio channel statistics. To measure the sensitivity of the decision process at the SNR fluctuations, Fig.~\ref{fig:convergence_vs_SNR} considers a fixed number of slices (i.e., $3$) deployed in a system characterized by average channel statistics with an increasing variance. In every scenario, the average (per slice) channel realization is derived from a Rayleigh distribution characterized by a scale parameter $\tau = \{0.1, 0.2,0.3,0.4\}$, respectively. This introduces an increasing level of variability in the SNR distribution according to the formula $\textit{Var} = \frac{4-\pi}{2}\tau^2$, as depicted in the plots of the central column.
On the left-hand side of the same picture, it can be noticed how higher SNR variability has very limited impact on the decision steps. This feature is inherited by the Markov Chain model described in Section~\ref{sect:MC}. In particular, provided that the slice requirements fit within the admissibility region of the system, a higher SNR variability will simply map into a wider excursion over the Markov chain steady states without affecting the final reward of the same arm.

Finally, on the right-hand side of the picture, we depict the empirical CDFs of the overall latency occurred per slice. 
In (almost) static channel conditions, slices' latency distribution suffer from having poor channel conditions, which are barely sufficient to support requested data volumes. In this context, slices with less stringent delay requirements, namely the MTC and eMBB, are lightly penalized to meet the expected latency threshold w.r.t. the URLLC one. When increasing the channel variability, the average channel conditions improve easing the allocation resource task thus favouring the satisfaction of overall latency requirements.
}

 \begin{figure*}[t!]
	\centering
	\begin{subfigure}[b]{.33\textwidth}
		\centering
      \includegraphics[trim = 14cm 0cm 1cm 1cm, clip, width=\linewidth ]{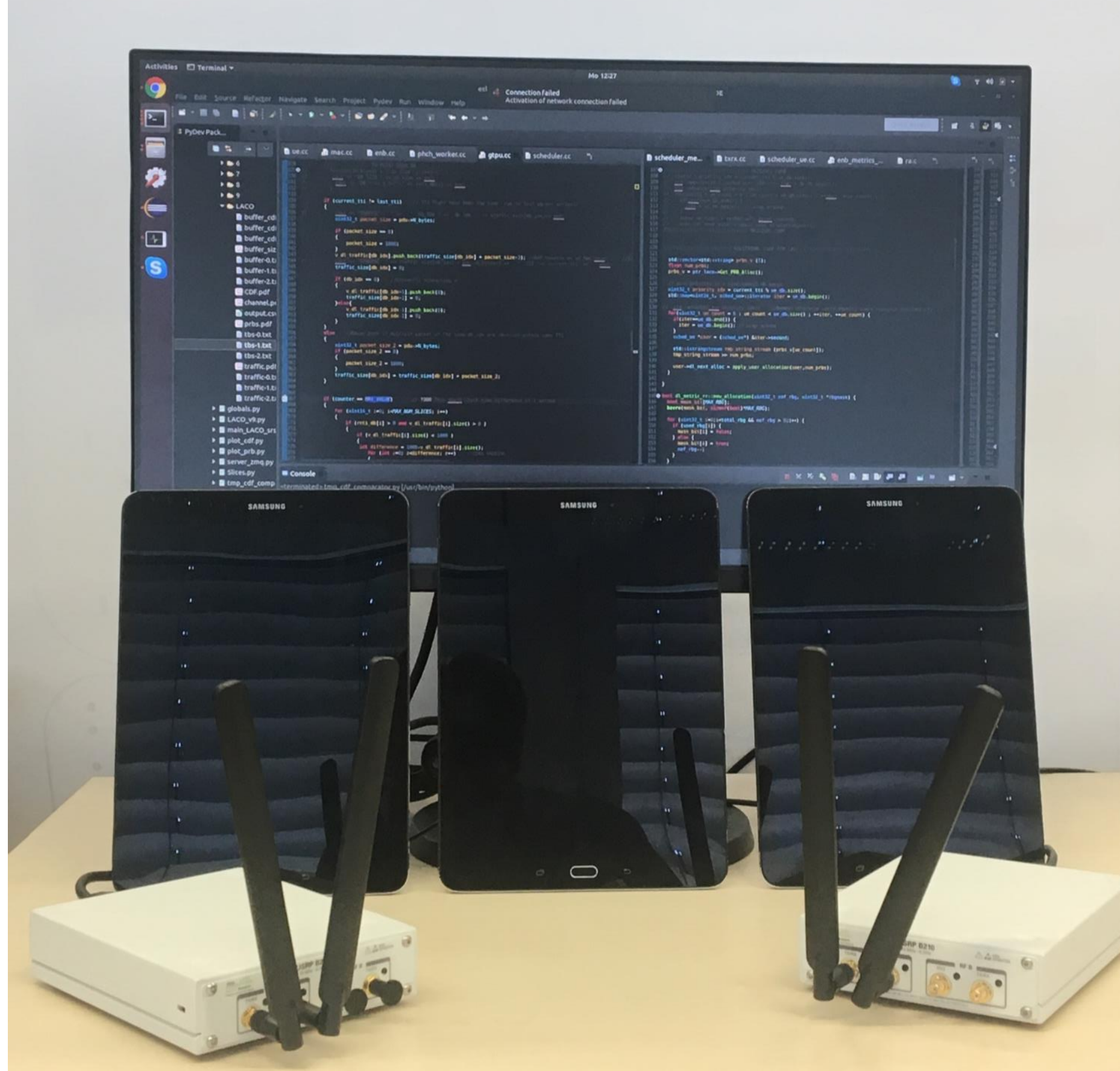}
      \caption{\small Experimental setup.}
      \label{fig:testbed}
    \end{subfigure}%
    \begin{subfigure}[b]{.33\textwidth}
    	\centering
      \includegraphics[trim = 17cm 1.5cm 1.5cm 1.2cm, clip, width=\linewidth ]{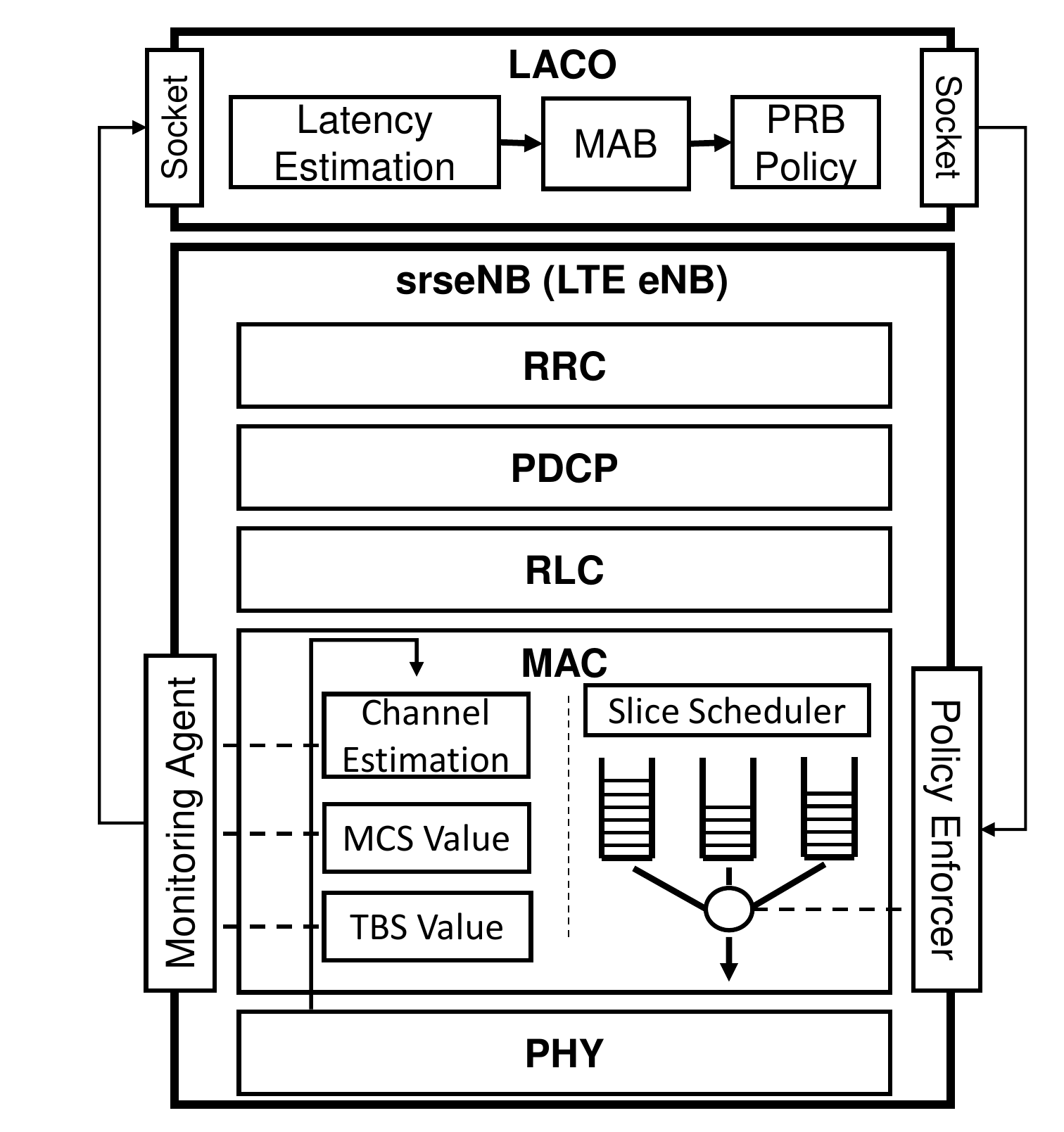}
      \caption{\small Architecture overview.}
      \label{fig:architecture}
      \end{subfigure}%
    \begin{subfigure}[b]{.33\textwidth}
    	\centering
      \includegraphics[trim = 0cm 0cm 0cm 0cm, clip, width=\linewidth ]{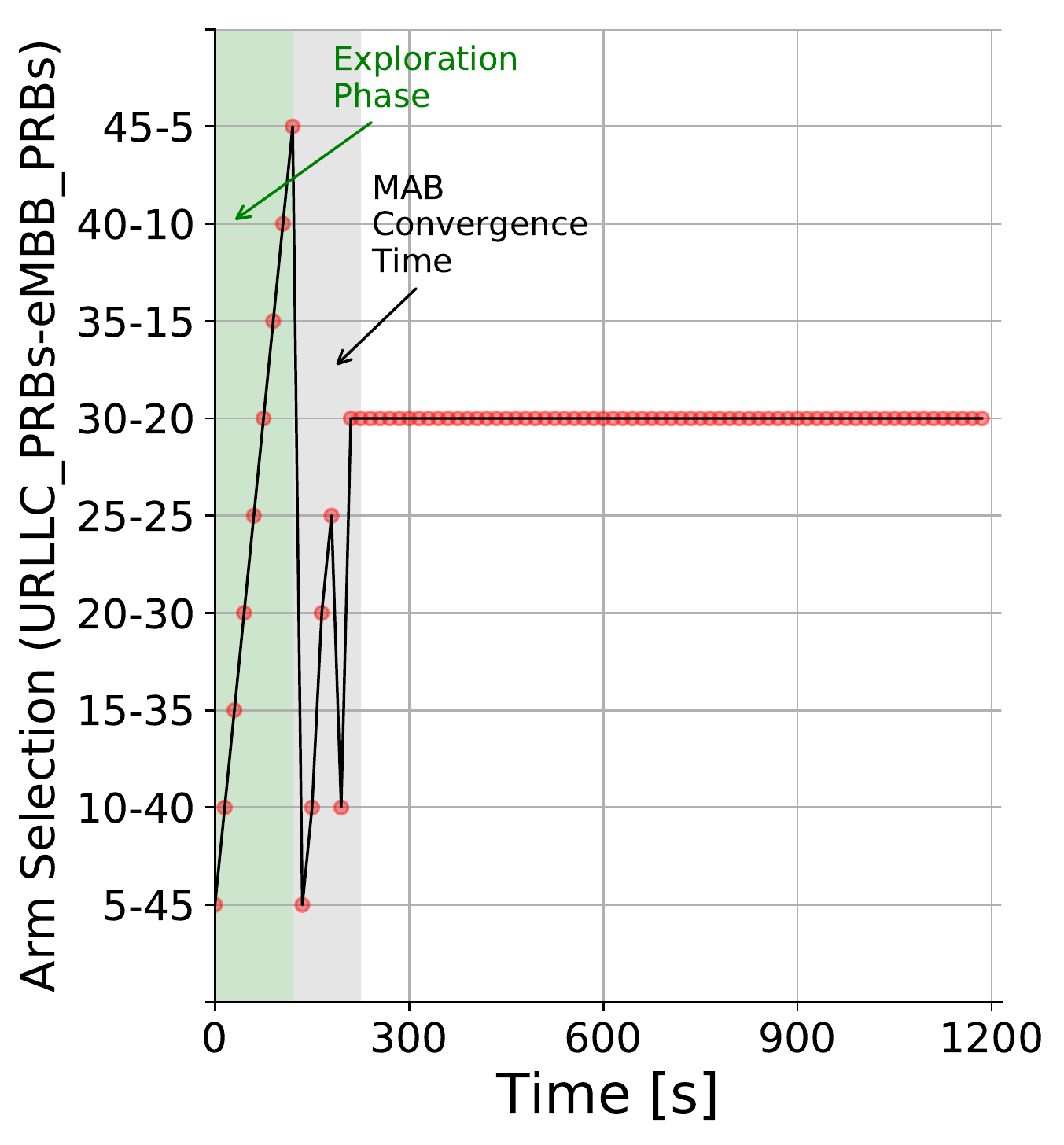}
      \caption{\small Arm selection.}
      \label{fig:arm_selection}
      \end{subfigure}%
      \caption{\small Experimental setup (a); Architecture overview (b); Arm selection over time (c).}
	\label{fig:POC_results}
	\vspace{-4mm}      
\end{figure*}

\section{Experimental Proof Of Concept}
\label{sect:poc}

In order to illustrate, validate and analyze the performance of our \name{} solution, we developed it as a standalone software module running on top of an open source platform that implements the LTE protocol stack, namely {\ttfamily srsLTE}~\cite{SRSLTE}, attached to a USRP\footnote{USRP B210 from National Instruments/Ettus Research (\url{https://www.ettus.com/all-products/UB210-KIT/}).} Software-Defined Radio (SDR) device as radio front-end. 
Our testbed is depicted in Fig.~\ref{fig:testbed} and consists of one LTE eNB (a modification of 
{\ttfamily srseNB}) and commercial Android tablets\footnote{Samsung Galaxy Tab S2 \url{ (https://www.samsung.com/de/tablets/galaxy-tab-s2-9-7-t813/SM-T813NZKEDBT/).}} as UEs. Any single UE emulates the aggregated traffic of multiple UEs within one slice. We use {\ttfamily mgen}\footnote{mgen (\url{https://www.nrl.navy.mil/itd/ncs/products/mgen}).} to generate different downlink traffic patterns.
Due to our LTE spectrum testing license restrictions, we use $10$~MHz bandwidth in LTE band $7$ and use SISO configuration for simplicity. This renders a maximum capacity of $\sim\!36$~Mb/s with highest SNR. Finally, in accordance with the findings described in Section~\ref{sec:MAB_behaviour}, we set the minimum PRB allocation value at $10\%$ of the overall availability.

\subsection{Implementation}
The architecture of our software implementation and \name{}'s interfaces with {\ttfamily srseNB} are depicted in Fig.~\ref{fig:architecture}. \name{} interacts with the eNB's Medium Access Control (MAC) layer to implement two key features:
\begin{itemize}
    \item {\bf Monitoring agent}. This feeds \name{} with real-time SNR reports generated by the physical (PHY) layer from feedback received from the UEs, the selected MCSs and corresponding transport block size (TBS) value used to encode information at the MAC layer, and other traffic statistics such as packet size and arrival times;
    \item {\bf Policy Enforcer}. This allows \name{} to dynamically enforce the PRB allocation policies calculated by our MAB model, as described in Section~\ref{sect:mabmodel}.
\end{itemize}

The main feature of our implementation is the possibility to
collect, with TTI granularity, the traffic arrival rate and the TBS values to be used in each transmission frame. This information, together with the scheduling buffer size and data arrival times, is essential to compute the latency experienced by the different slices running in the system.

The different metrics are collected in a time series database, namely {\ttfamily InfluxDB}, and periodically reported to \name{} which constructs a virtual queue (one per slice) to track the dynamics of packets arriving at the eNB, from their entrance into the scheduling buffer to their transmission.
This approach is particularly useful as Internet Protocol (IP) packets are multiplexed while advancing the transmission path in the eNB, complicating the computation of slice latencies by external modules. Our approach aims to characterize the PRB allocation policy currently enforced into the system. In case of constant traffic and low latency requirements for example, poor channel conditions will result in lower TBS values and a sudden increase of the virtual queues size. Such event directly maps into an additional delay suffered by IP packets at the Radio Link Control (RLC) layer.
Note that higher packet rates also lead to larger waiting times, which might result in exceeding slices SLAs boundaries. In such cases, the violation of pre-defined SLA latency boundaries triggers the DTMC model described in Section~\ref{sect:MC} to a \emph{delay state} and the selected PRB allocation policy is assigned with a lower reward value.
Conversely, in a stable system where serving rate and packet arrival rate are balanced, the size of the virtual queues get smaller and the DTMC model is mostly characterized by \emph{non-delay states}.


\begin{figure*}[t!]
	\centering
	\begin{subfigure}[b]{.5\textwidth}
	    \centering
        \includegraphics[trim = 0cm 0cm 0cm 0cm, clip, width=\linewidth ]{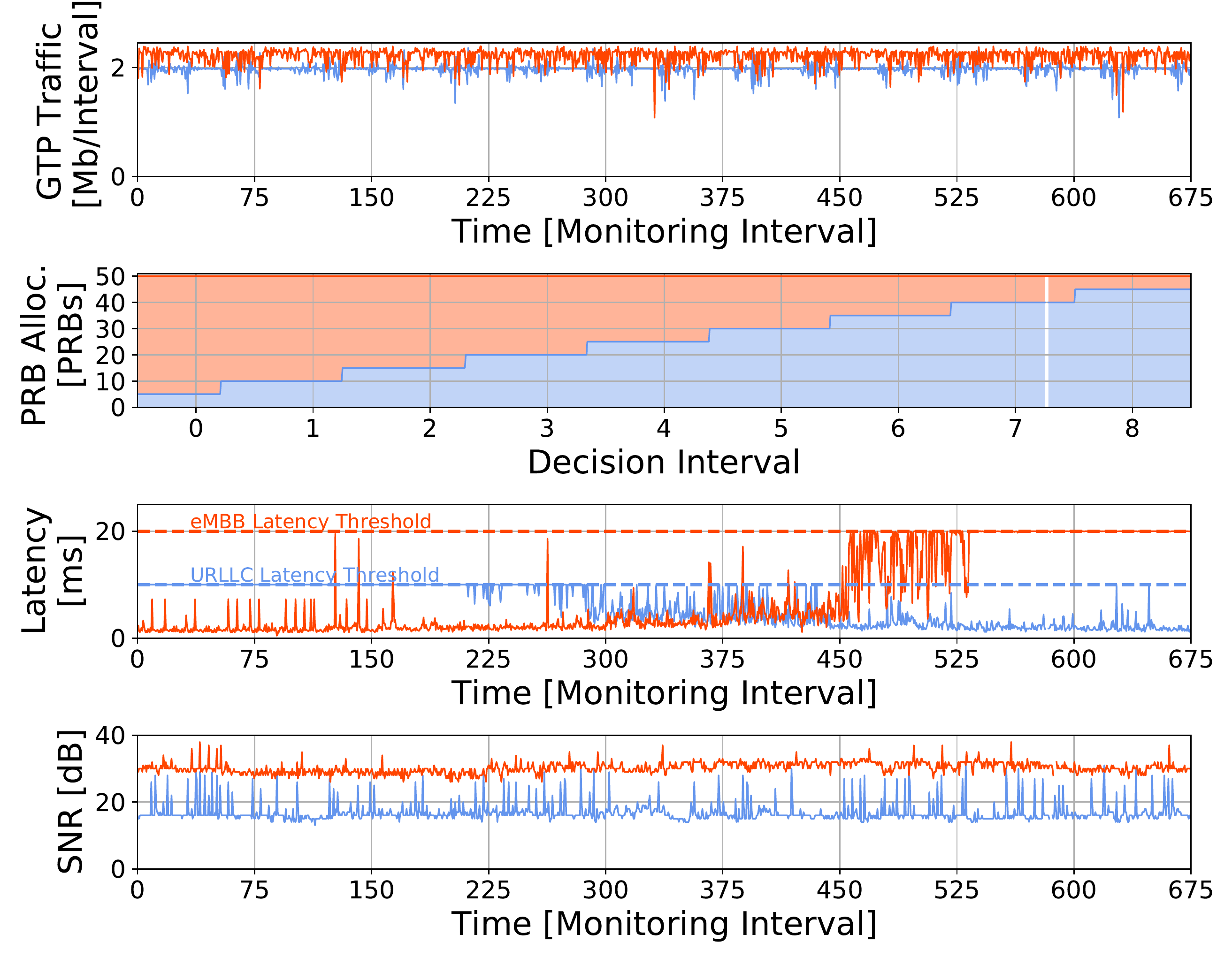}
        \caption{\small System dynamics during MAB discovery phase.}
        \label{fig:discovery_phase}
    \end{subfigure}%
    \begin{subfigure}[b]{.5\textwidth}
    \centering
      \includegraphics[trim = 0cm 0cm 0cm 0cm, clip, width=\linewidth ]{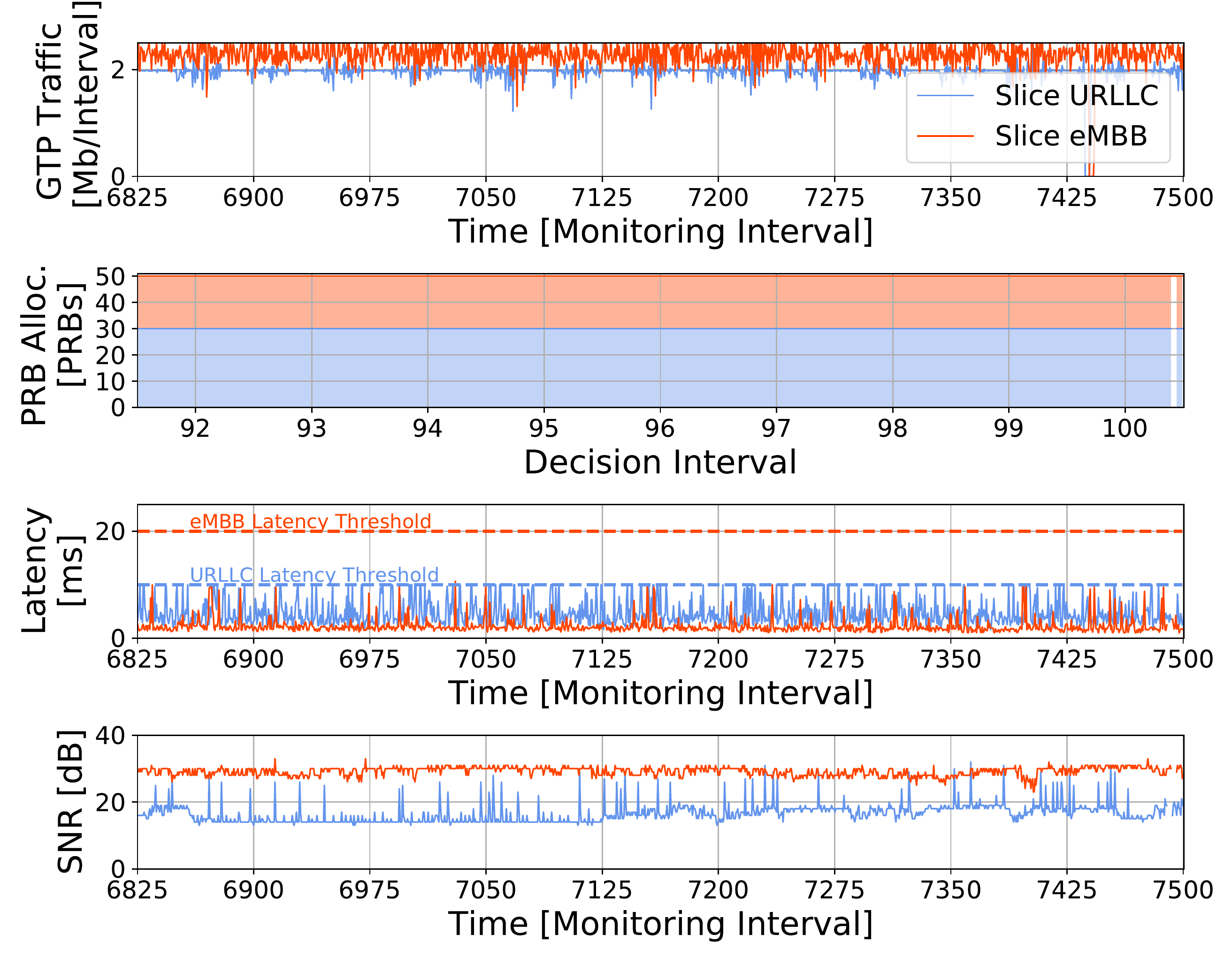}
    \caption{\small System dynamics at convergence.}
    \label{fig:final_phase}
      \end{subfigure}%
      \caption{\small Comparison of system dynamics during a) discovery phase and b) MAB convergence.}
	\label{fig:MAB_Evolution}
	\vspace{-4mm}      
\end{figure*}

\subsection{Experimental results}
We consider a scenario accounting for two slices characterized by the following requirements. The first slice (labelled Ultra Reliable Low Latency Communications or URLLC) demands $\Delta_{\textit{URLLC}}=10$~ms communication delay and is characterized by a constant bit rate equal to $9.6$~Mb/s. The second slice (labelled enhanced Mobile Broaband or eMBB) is characterized by a constant throughput equal to $11.2$~Mb/s with a more relaxed latency requirements $\Delta_{\textit{eMBB}}=20$~ms.
We set \name{}'s decision interval to $15$ seconds and let our experiment run over the downlink direction for $100$ decision intervals. Fig.~\ref{fig:arm_selection} shows the evolution of the PRB allocation configuration decisions taken by \name{} over this time span and how fast the convergence to a suitable layout is achieved. 
The monitoring information about incoming traffic at GTP level collected during the experiment are depicted in the upper plots of Figs.~\ref{fig:discovery_phase} and \ref{fig:final_phase}. It should be noticed that these values represent aggregated values (sum) over monitoring intervals of $200$ms. Latency and SNR information are depicted in the third and fourth plots of each figure. In this case, we use maximum and average as aggregation functions, respectively.

\begin{figure*}[t!]
	\centering
	\begin{subfigure}[b]{.5\columnwidth}
	    \centering
        \includegraphics[trim = 0cm 0cm 0cm 0cm, clip,     width=0.95\linewidth]{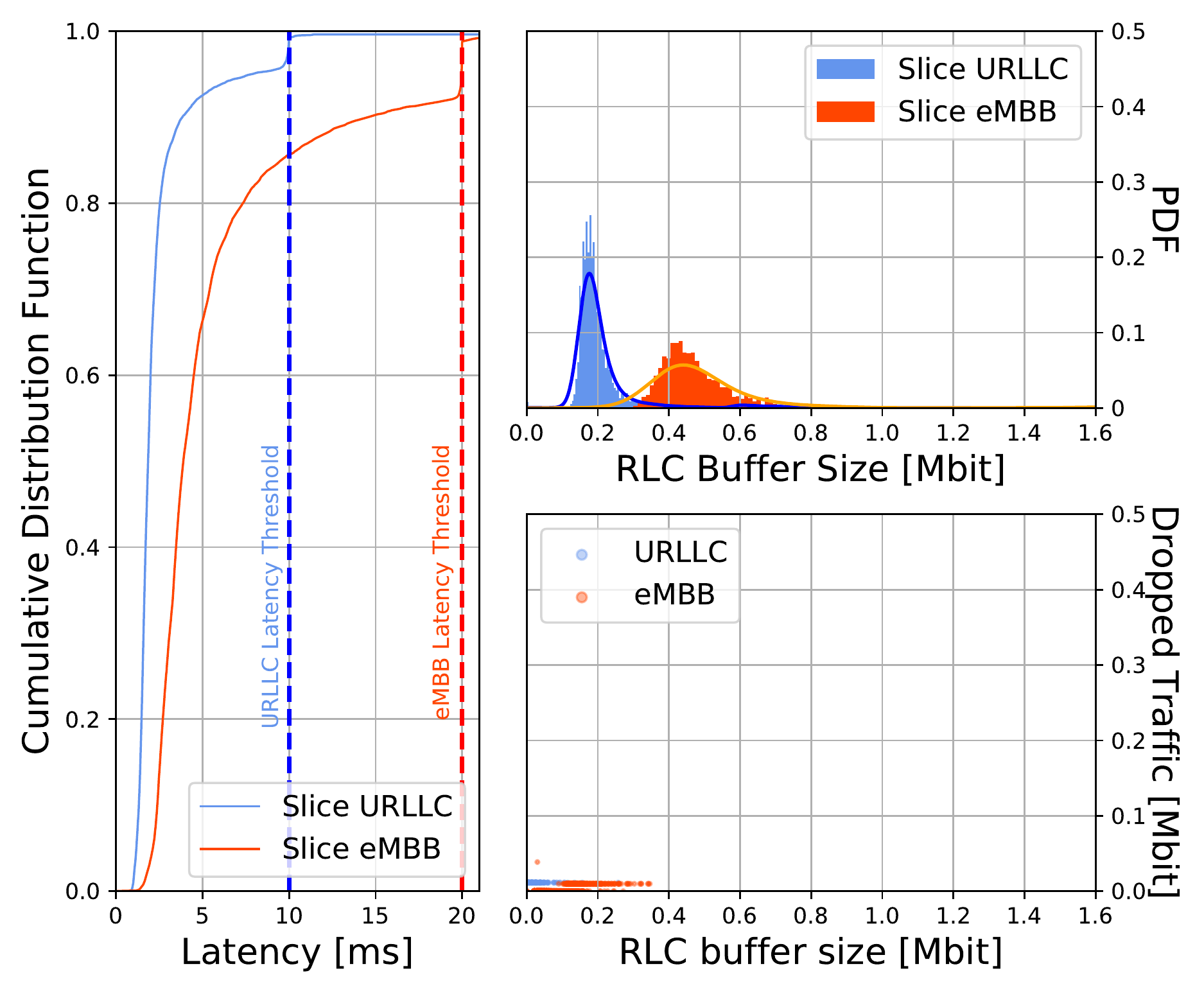}
        \vspace{-2mm}
        \caption{\small \name{}}
        \label{fig:laco_comparison}
    \end{subfigure}%
    \begin{subfigure}[b]{.5\columnwidth}
	    \centering
        \includegraphics[trim = 0cm 0cm 0cm 0cm, clip,     width=0.95\linewidth]{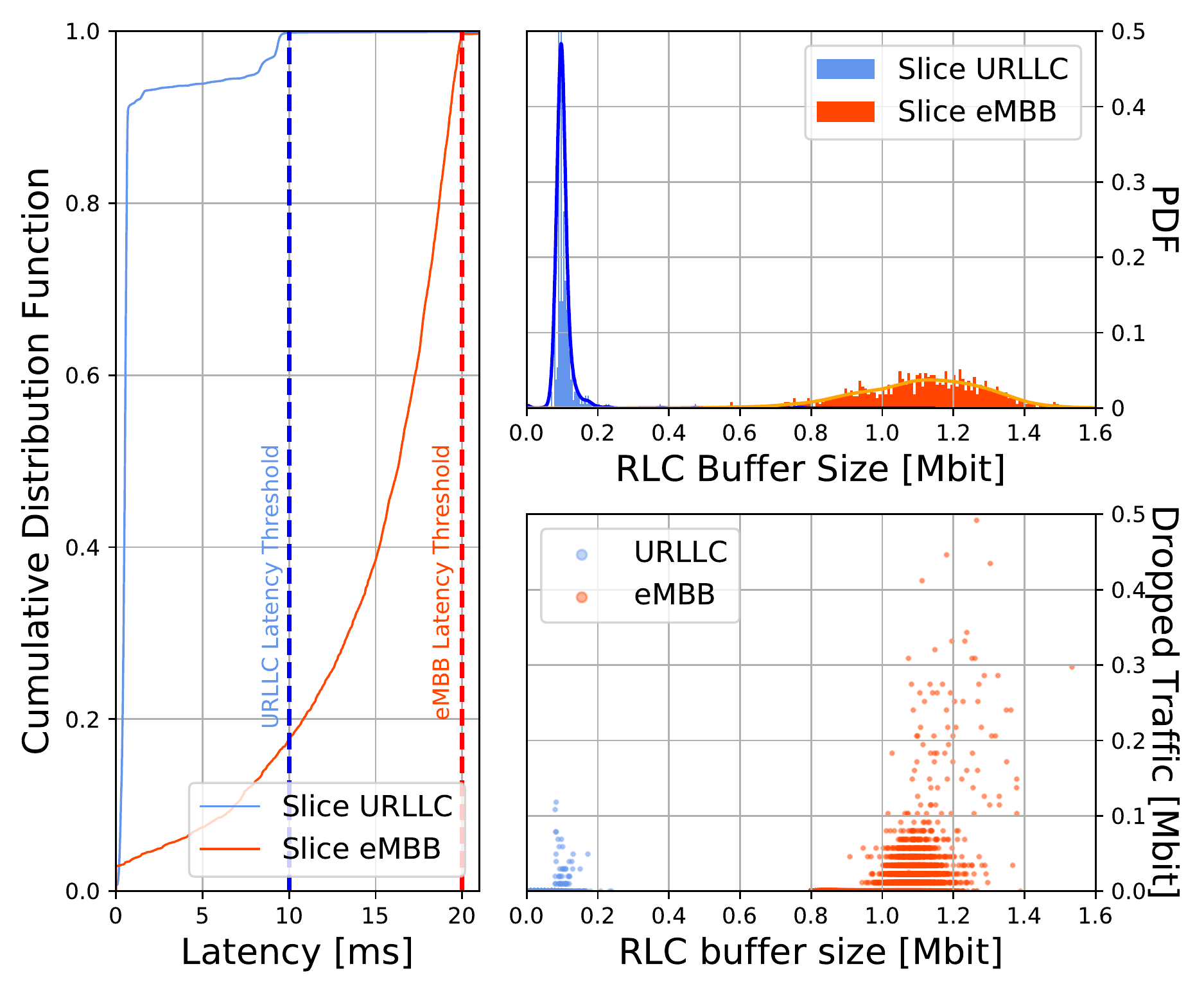}
        \vspace{-2mm}
        \caption{\small Round Robin ($RR$)}
        \label{fig:rr_comparison}
    \end{subfigure}%
    \caption{\small Evaluation of different performance metrics for different scenarios.}
    \vspace{-4mm}
    \label{fig:poc_results}
\end{figure*}

As described in Section~\ref{sect:mabmodel}, during the starting procedure the MAB algorithm explores all available arms with the aim of collecting an initial feedback on the system dynamics. Fig.~\ref{fig:discovery_phase} 
depicts the effects of these sequential choices on the latency experienced by the ongoing traffic flows. The initial steps drive the allocation of radio resources towards the eMBB slice thereby providing significant advantages in terms of experienced delay with respect to the URLLC one. In this phase, traffic coming from the URLLC might be dropped due to delay violation $\Delta_{\textit{URLLC}}$. The scenario changes after the $6$-th decision interval, when the agent selects the configuration ($35$-$15$). Given the current channel quality, that arm does satisfy the URLLC radio requirements but does not reserve enough radio resources for the eMBB slice, thus increasing the latency experienced by its users. Subsequent arm selections within decision intervals $7$ and $8$, further reduce the radio resources assigned to the eMBB slice thus leading the traffic to violate $\Delta_{\textit{eMBB}}$. The MAB agent collects this information and quickly converges to a satisfactory configuration. 
In Fig.~\ref{fig:final_phase}, we focus on the system dynamics once the convergence is achieved and clearly notice how both the latency requirements are satisfied. Interestingly, despite similar traffic requirements, the algorithm selects the configuration ($30$-$20$), which assigns more resources to the first slice. This is justified by the lower SNR value experienced by such a slice during the experiment, as depicted in the bottom plot of Figs.~\ref{fig:discovery_phase} and \ref{fig:final_phase}. The URLLC slice thus requires more PRBs to compensate for the lower MCS used during the communication and successfully meets the latency requirements.
For illustration purposes, we select a vanilla PRB allocation policy, namely round-robin~(RR), as a generic non-latency-aware benchmark and compare the performance of the two schemes running in the same scenario. The results of our experiments are summarized in Fig.~\ref{fig:poc_results}, where both plots depict the empirical CDF of the latency, the RLC buffer density and the dropping rate incurred by each slice for the two allocation schemes. 

The performances of the system when \name{} is in place are depicted on the left-hand side picture, whereas the right-hand side shows the results of the RR-based slice scheduling scheme. In both plots, the URLLC slice is shown in blue and the eMBB one in orange. Based on these results, we can observe that \name{} successfully meets both slices latency requirements. This is achieved by providing the required resources to the URLLC and eMBB slices (Fig.~\ref{fig:arm_selection}) according to their different latency needs. This results in the URLLC slice allowing SLA latency requirements ($\le10$ms) at a very low average latency cost increase for the eMBB slice. In our experiments, very few traffic ($\sim$ $2\%$) experienced a latency above the $10$~ms target of URLLC when using \name{}, in contrast to $\sim$ $10\%$ experienced with RR. Despite of negligible impact, note that by our design choice parts of fragmented packets are sent even if above the latency threshold to avoid long HARQ based retransmission procedures~\cite{TS36.321}, which may negatively affect the slice performance. Moreover, we wish to highlight that for \name{} the amount of violations due to the exploration and convergence period could be significantly reduced if desired by introducing a policy aimed at minimizing such cases.
The performance gap further increases when comparing the eMBB results. Given that $RR$ sequentially allocates resources to the URLLC slice and, when the buffer is empty, to eMBB, it consistently favours the URLLC slice over the eMBB one. Thus, despite the higher channel quality condition experienced by the eMBB slice, in every scheduling period the resource availability for the eMBB slice is highly reduced. This provides a better performances for URLLC traffic, but at a significant degradation cost for eMBB users, as confirmed by Fig.~\ref{fig:rr_comparison} (bottom-right), which depicts the amount of traffic dropped during the experiment.
The latency performance is strongly related with the traffic queue waiting in the transmission buffers. For this reason, the two figures depict the buffer size density distribution obtained during the experiments. It is clear from the comparison how different PRB allocation schemes affect the transmission buffer size at RLC layer. In the \name{} case, they are generally lightly loaded, finally providing shorter serving time for incoming packets. In the $RR$ scenario however, the eMBB traffic suffers higher congestion, which leads to augment packet's waiting time, and consequently increases the rate of latency violations.


\section{Related Work}
\label{sect:related}

The RAN design problem has always been at the forefront of the mobile operators and a vast amount of research has been devoted to novel RAN architectures~\cite{Rostami, fluidran-jsac} and efficient radio resource schedulers~\cite{doro2017, lasr}. Recently, network slicing has been proposed as a new means for mobile operators to deploy isolated network services owned by different customers over a common physical infrastructure. However, as highlighted in~\cite{da2016impact}, RAN needs additional functionalities to fully exploit SDN and NFV principles, specially in the partition and isolation of radio resources. The authors of~\cite{foukas_orion} focus on efficient sharing of the RAN resources and proposed a RAN slicing solution that performs adaptive provisioning and isolation of radio slices. Their work is based on dynamic virtualization of base station resources, which gives tenants the ability to independently manipulate each slice. Although the proposed architecture may guarantee isolation through different control planes, no mechanism is in place to ensure the satisfaction of delay requirements. \cite{mobicom2018_MGFBC} provides an empirical study of resource management efficiency in slicing-enabled networks through real data collected from an operational mobile network, considering different kinds of resources and including radio access, transport and core of the network. Similarly, the authors of~\cite{Chang2018} formulate an optimization framework to deal with resource partitioning problem, where inter-slice isolation is assured through a virtualized layer that decouples the reservation choice from the physical resource availability and proposing different abstraction types of radio resource sharing. \change{In~\cite{Guo} the authors present an Earliest Deadline First (EDF) scheduling approach in the context of network slicing. Differently from us, their approach works on a single MAC scheduler and assumes for every TTI a complex fine-tuning of the quota of resources to be assigned to each slice, thus limiting the implementation of dedicated intra-scheduling solutions.
}

The exploration-vs-exploitation trade-off, typical of Multi-Armed Bandit (MAB) problems is particularly suited to problems that require sequential decision-making. For this reason, a wide set of variations from the classical MAB model has been proposed in the literature~\cite{survery_mab, Mahajan2008}, together with novel algorithms to address them~\cite{Algorithms2000}. In this regard, the work of~\cite{OnlineMAB-Tekin} investigates the MAB problems in case of Markovian reward distribution, where arms change their state in a two-state Markovian fashion. The authors addressed the problem assuming that the Markov chain evolves only when the arm is played, showing that the proposed sample mean-based index policy achieves regret performances comparable to legacy UCB algorithm.
The authors of~\cite{Agrawal2017} performed a complete regret analysis of the TS algorithm, generalizing the original formulation to distributions other than the Beta distribution. 
The MAB framework is also applied in~\cite{Proutiere_sigmetrics18} to deal with rate adaptation problem in $802.11$-like wireless systems. The authors demonstrate that exploiting additional observations significantly improve the system performance. Similarly, \cite{Stahlbuhk2018_mobihoc} deals with scheduling transmissions in presence of unknown channel statistics. The proposed algorithm learns the channels' transmission rates while simultaneously exploiting previous observations to obtain higher throughput. This led to the design of a queue-length-based scheduling policy using the channel learning algorithm as a component in time-varying environment.
The authors of~\cite{Hill2017} presented an algorithm for multivariate optimization on large decision spaces based on an innovative approach combining hill climbing optimization and Thompson sampling. While the scalability of their algorithm has been proven through exhaustive simulations, the framework lacks a complete analysis of regret bounds aimed at demonstrating the impact of hill climbing in combinatorial decision making.
Finally, similar to us, \cite{Besbes2014} deals with an MAB formulation where the reward distributions are characterized by temporal uncertainties. Interestingly, they were able to mathematically capture, in terms of reward, the added complexity embedded in the non-stationarity feature when compared to the legacy framework.

The key novelty of \name{} relies on the exploitation of (abstract) information of the underlying system structure to expedite solutions. Conversely, prior works are blind to such type of information and need to spend substantial time exploring very bad decisions before achieving it.


\section{Conclusions}
\label{sect:concl}

Major efforts in the design of next-generation mobile systems pivot around network slicing and (mobile edge) low-latency services. This paper aims to bridge the gap between them both by designing \name, a RAN-specific network slice orchestrator that considers network slice requests with \emph{strict latency requirements}. Despite the efforts devoted by 5G researchers and engineers to network slicing, to the best of our knowledge, this is the first radio slicing mechanism that provides \emph{formal delay guarantees}. 
To make network slicing decisions in environments with varying wireless channel quality and user demands, \name{} builds on a learning Multi-Armed Bandit (MAB) method that is \emph{model-aware} as opposed to classic MAB approaches that are blind to information regarding the underlying system. In addition, we exploit information from the system model to expedite the exploration-vs-exploitation process. Our results derived from an implementation with off-the-shelf hardware show that \name{} is able to guarantee strict slice latency requirements at affordable computational costs.

\section*{Acknowledgment}
The research leading to these results has been partially supported by the H$2020$ MonB5G Project under grant agreement number $871780$.

\bibliographystyle{IEEEtran}
\bibliography{main}

\end{document}